\documentclass[twoside]{article}
\usepackage{amsfonts, amsthm, amssymb}
\usepackage[mathcal]{eucal}

\newtheorem{theorem}{Theorem}[section]
\newtheorem{lemma}[theorem]{Lemma}
\newtheorem{prop}[theorem]{Proposition}
\newtheorem{cor}[theorem]{Corollary}
\newtheorem{remark}[theorem]{Remark}
\newtheorem{definition}{Definition}[section]

\def\l{\lambda}
\def\a{\alpha}
\def\b{\beta}

\def\g{\gamma}

\def\e{\varepsilon}

\def\z{\zeta}

\def\R{\mathbb{R}}
\def\C{\mathbb{C}}
\def\N{\mathbb{N}}
\def\Z{\mathbb{Z}}

\def\Q{\mathbb{Q}}

\def\Hab{H_{\alpha, \beta}}
\def\H0b{H_{0, \beta}}
\def\Ha0{H_{\alpha, 0}}

\def\Qab{Q_{\alpha, \beta}}
\def\Pab{P_{\alpha, \beta}}
\def\gab{\g(\a,\b)}
\def\hvt{\tilde{H}_V}

\def\mbf2{\mathbf{2}}

\def\sk{\sin\frac{k\pi}{N}}
\def\skk{\sin\frac{k'\pi}{N}}
\def\skkk{\sin\frac{k''\pi}{N}}

\def\sl{\sin\frac{l\pi}{N}}
\def\sm{\sin\frac{m\pi}{N}}
\def\sn{\sin\frac{n\pi}{N}}

\def\1N1{1 \leq k \leq N-1}

\begin{document}

\title{Normal form for odd periodic FPU chains}
\author{Andreas Henrici\footnote{Supported in part by the Swiss National Science Foundation} \and Thomas Kappeler\footnote{Supported in part by the Swiss National Science Foundation, the programme SPECT and the European Community through the FP6 Marie Curie RTN ENIGMA (MRTN-CT-2004-5652)}}

\maketitle

\begin{abstract}
In this paper we prove that near the equilibirum position any periodic FPU chain with an odd number of particles admits a Birkhoff normal form up to order $4$, and we obtain an explicit formula of the Hessian of its Hamiltonian at the fixed point.
\end{abstract}

\section{Introduction} \label{introduction}

In this paper we consider FPU chains with $N$ particles of equal mass (normalized to be one). Such chains have been introduced by Fermi, Pasta, and Ulam \cite{fpu}, as models to test numerically the phenomenon of thermalization as the number of particles gets larger and larger. A FPU chain consists of a string of particles moving on the line or the circle interacting only with their nearest neighbors through nonlinear springs. Its Hamiltonian is given by
\begin{equation} \label{hvgendef}
  H_V = \frac{1}{2} \sum_{n=1}^N p_n^2 + \sum_{n=1}^N V(q_{n+1} - q_n),
\end{equation}
where $V: \R \to \R$ is a smooth potential. The corresponding Hamiltonian equations read $(1 \leq n \leq N)$
\setlength\arraycolsep{2pt}{\begin{eqnarray*}
  \dot{q}_n & = & \partial_{p_n} H_V = p_n, \\
  \dot{p}_n & = & -\partial_{q_n} H_V = V'(q_n - q_{n+1}) - V'(q_{n-1} - q_n).
\end{eqnarray*}}
Here $q_n$ denotes the position of the $n$'th particle relative to its equilibrium position, $p_n$ is its momentum, and throughout this paper we assume periodic boundary conditions
\begin{displaymath}
  (q_{i+N}, p_{i+N}) = (q_i, p_i) \quad \forall i \in \{ 0,1 \}.
\end{displaymath}

Without loss of generality, the potential $V: \R \to \R$ is assumed to have a Taylor expansion at $0$ of the form
\begin{equation} \label{potentialdef}
  V(x) = \kappa \left( \frac{1}{2} x^2 + \frac{\a}{3!} x^3 + \frac{\b}{4!} x^4 + \ldots \right),
\end{equation}
where $\kappa$ is the (linear) spring constant normalized to be $1$ and $\a, \b \in \R$ are parameters measuring the strength of the nonlinear interaction. Substituting the expression (\ref{potentialdef}) for $V$ into (\ref{hvgendef}), the corresponding expansion of $H_V$ is given by
\begin{equation} \label{hvspecialdef}
  H_V = \frac{1}{2} \sum_{n=1}^N p_n^2 + \frac{1}{2} \sum_{n=1}^N (q_n - q_{n+1})^2 + \frac{\a}{3!} \sum_{n=1}^N (q_n - q_{n+1})^3 + \frac{\b}{4!} \sum_{n=1}^N (q_n - q_{n+1})^4 + \ldots.
\end{equation}

For any FPU chain, the total momentum $P = \frac{1}{N} \sum_{n=1}^N p_n$ is an integral of motion,
\begin{displaymath}
  \dot{P} = \frac{1}{N} \sum_{n=1}^N \dot{p}_n = \frac{1}{N} \sum_{n=1}^N (V'(q_n - q_{n+1}) - V'(q_{n-1} - q_n)) = 0,
\end{displaymath}
and therefore the center of mass $Q = \frac{1}{N} \sum_{n=1}^N q_n$ evolves with constant velocity. Hence any FPU chain can be reduced to a family of Hamiltonian systems of $2N-2$ degrees of freedom, parametrized by the vector of initial conditions $(Q,P) \in \R^2$ with Hamiltonian independent of $Q$. In particular, for $N=2$ any FPU chain is integrable. Further note that for any vector $(Q,P) \in \R^2$, the origin in $\R^{N-2}$ is an equilibrium point of the reduced system. The momentum of such an equilibrium point is given by the constant vector $(p_1, \ldots, p_N) = P \, (1, \ldots, 1)$.

Introduce the functions $I = (I_k)_{\1N1}$, $J = (J_k)_{\1N1}$, and $M = (M_k)_{\1N1}$ defined on $\R^{2N-2}$ with values in $\R^{N-1}$ given by
\begin{equation} \label{actiondef}
  I_k = \frac{1}{2} (x_k^2 + y_k^2); \quad J_k = \frac{1}{2} (x_k x_{N\!-\!k} + y_k y_{N\!-\!k}); \quad M_k = \frac{1}{2} (x_k y_{N\!-\!k} - x_{N\!-\!k} y_k).
\end{equation}
Further define the function $\Hab: \R^{N-1} \to \R$, given by
\begin{equation} \label{bnfintrotheorem}
  \Hab(I) := 2 \sum_{k=1}^{N-1} \sin \frac{k \pi}{N} I_k + \frac{1}{4N} \sum_{k=1}^{N-1} c_k I_k^2 + \frac{\b - \a^2}{2N} \!\!\!\!\!\!\!\! \sum_{l \neq m \atop 1 \leq l,m \leq N-1} \!\!\!\!\!\!\!\! \sin \frac{l \pi}{N} \sin \frac{m \pi}{N} I_l I_m,
\end{equation}
where $c_k \equiv c_k(\a,\b) := \a^2 + (\b-\a^2) \sin^2 \frac{k \pi}{N}$.

The main result of this paper is
\begin{theorem} \label{bnffputheorem}
If $N \geq 3$ is odd, any periodic FPU chain admits a Birkhoff normal form of order $4$ (included). More precisely, for any odd $N \geq 3$ there are canonical coordinates $(x_k, y_k)_{\1N1}$ so that the Hamiltonian of \emph{any} FPU chain, when expressed in these coordinates, takes the form
\begin{displaymath}
\frac{N P^2}{2} + \Hab(I) + O(|(x,y)|^5)
\end{displaymath}
with $\Hab(I)$ given by (\ref{bnfintrotheorem}).
\end{theorem}

\begin{cor}
Near the equilibrium state any FPU chain with an odd number $N$ of particles can be approximated up to order $4$ relative to its center of mass coordinates by an integrable system of $N-1$ harmonic oscillators which are \emph{coupled} at fourth order.
\end{cor}

Denote by $\Qab$ the Hessian of $\Hab(I)$ at $I=0$. Note that $\Qab$ is an $(N-1) \times (N-1)$ matrix which only depends on the paramters $\a$ and $\b$. For the following result we do not have to assume that $N$ is odd.

\begin{theorem} \label{bnfproperties}
  \begin{itemize}
  \item[(i)] For any given $\a \in \R \setminus \{ 0 \}$, $\det (\Qab)$ is a polynomial in $\b$ of degree $N-1$ and has $N-1$ real zeroes (counted with multiplicities). When listed in increasing order the zeroes $\b_k = \b_k(\a) (\1N1)$ satisfy
\begin{displaymath}
  0 < \b_1 < \a^2 < \b_2 \leq \ldots \leq \b_{N-1}  
\end{displaymath}
and contain the $\llcorner \frac{N-1}{2} \lrcorner$ distinct numbers
\begin{displaymath}
  \a^2 \left( 1 + (\sin^2 \frac{k \pi}{N})^{-1} \right) \quad (1 \leq k \leq \llcorner \frac{N-1}{2} \lrcorner).
\end{displaymath}
Moreover, index$(\Qab)$, defined as the number of negative eigenvalues of $\Qab$, is given by
\begin{displaymath}
  \textrm{index} \, (\Qab) = \left\{  \begin{array}{ll}
1 & \quad \textrm{for } \b < \b_1 \\
0 & \quad \textrm{for } \b_1 < \b < \b_2 \\
N-2 & \quad \textrm{for } \b > \b_{N-1}
\end{array} \right.
\end{displaymath}
\item[(ii)] For $\a = 0$, $\det(Q_{0, \b})$ is a polynomial in $\b$ of degree $N-1$, and $\b = 0$ is the only zero of $\det(Q_{0, \b})$. It has multiplicity $N-1$, and the index of $Q_{0, \b}$ is given by
\begin{displaymath}
  \textrm{index} \, (Q_{0, \b}) = \left\{  \begin{array}{ll}
1 & \quad \textrm{for } \b < 0 \\
N-2 & \quad \textrm{for } \b > 0
\end{array} \right.
\end{displaymath}
\end{itemize}
\end{theorem}

FPU chains with an even number of particles typically (i.e. if $\b \neq \a^2$) do not admit a Birkhoff normal form up to order $4$ due to resonances. Our analysis of odd FPU chains leads in the case of even FPU chains to a \emph{resonant} Birkhoff normal form up to order $4$ which we use in subsequent work \cite{ahtk5} to show that their Hamiltonians truncated at fourth order are nevertheless integrable systems in the sense of Liouville.

Recall the definiton (\ref{actiondef}) of the functions $J$ and $M$. Let
\begin{equation} \label{rabjmdef}
	R_{\a, \b}(J,M) := \frac{\b-\a^2}{4N} \left( R(J,M) + R_\frac{N}{4}(J,M) \right)
\end{equation}
where
\begin{displaymath}
R(J,M) = 4 \sum_{1 \leq k < \frac{N}{4}} \!\! \sin \frac{2 k \pi}{N} \left( J_k J_{\frac{N}{2} - k} - M_k M_{\frac{N}{2} - k} \right)
\end{displaymath}
and
\begin{displaymath}
  R_\frac{N}{4}(J,M) = \left\{ \begin{array}{ll}
J_\frac{N}{4}^2 - M_\frac{N}{4}^2 & \quad \textrm{if } \frac{N}{4} \in \N \\
0 & \quad \textrm{otherwise.} \end{array} \right.
\end{displaymath}

\begin{theorem} \label{bnffputheoremeven}
If $N \! \geq \! 4$ is even, there are canonical coordinates $(x_k, y_k)_{1 \! \leq \! k \leq \! N-1}$ so that the Hamiltonian of \emph{any} FPU chain, when expressed in these coordinates, takes the form
\begin{displaymath}
\frac{N P^2}{2} + \Hab(I) - R_{\a, \b}(J,M) + O(|(x,y)|^5),
\end{displaymath}
where $\Hab(I)$ and $R_{\a, \b}(J,M)$ are given by (\ref{bnfintrotheorem}) and (\ref{rabjmdef}), respectively.
\end{theorem}

\begin{remark}
Theorem \ref{bnffputheoremeven} can be used to show that a version of Theorem \ref{bnffputheorem} holds for $N$-particle FPU chains ($N$ even or odd) considered with Dirichlet boundary conditions by embedding such systems into an invariant submanifold of a periodic system with $2N+2$ particles - see \cite{ahtk5} for details.
\end{remark}

\emph{Applications:} In the case where $N$ is odd, Theorems \ref{bnffputheorem} and \ref{bnfproperties} allow to apply for any given $\a \in \R$ the classical KAM theorem (see e.g. \cite{poeschelkam}) near the equilibrium point to the FPU chain with Hamiltonian $H_V$ for a real analytic potential $V = \frac{1}{2} x^2 + \frac{\a}{3!} x^3 + \frac{\b}{4!} x^4 + \ldots$ with $\b \in \R \setminus \{ \b_1(\a), \ldots, \b_{N-1}(\a) \}$. Moreover, as for any given $\a \in \R \setminus \{ 0 \}$, $\Qab$ is positive definite for $\b_1(\a) < \b < \b_2(\a)$, one can apply Nekhoroshev's theorem (see e.g. \cite{poeschelnekh2}) to the FPU chain with Hamiltonian $H_V$ for $V$ with such $\b$'s. These perturbation results confirm long standing conjectures - see e.g. \cite{beiz}.

\emph{Related work:} Theorem \ref{bnffputheorem} improves on earlier results of Rink \cite{rink01} and together with Theorem \ref{bnfproperties} solves all open problems stated in \cite{rink01} for $N$ odd. Instead of using symmetry properties of FPU chains employed in \cite{rink01} our approach has been shaped by our earlier work on the Toda lattice \cite{ahtk2, ahtk3}. The latter one, introduced by Toda \cite{toda} and extensively studied in the sequel, is a special FPU chain which is integrable. It turns out that (almost) the same canonical transformations which near the equilibrium bring the Toda lattice into Birkhoff normal form can be used for any FPU chain. Put in other words, the existence of the Birkhoff normal form stated in Theorem \ref{bnffputheorem} is, at least partially, a consequence of the fact that the family of FPU chains, parametrized by $\a, \b, \ldots$, contains an integrable system, namely the Toda lattice.

\emph{Outline:} In section \ref{bnftheory}, we review the notion of a Birkhoff normal form. We show Theorem \ref{bnffputheorem} in sections \ref{relcoord}-\ref{mainproof} and Theorem \ref{bnffputheoremeven} in section \ref{evenproof}, whereas Theorem \ref{bnfproperties} will be proved in section \ref{special}.

\emph{Acknowledgement:} It is a great pleasure to thank Yves Colin de Verdi\`ere and Percy Deift for valuable comments.

\section{Birkhoff normal form} \label{bnftheory}

Consider an isolated equilibrium of a Hamiltonian system on some $2n$-dimensio\-nal symplectic manifold, i.e. an isolated singular point of the Hamiltonian vector field. Neglecting an irrelevant additive constant, the Hamiltonian, when expressed in canonical coordinates $w = (q,p)$ near the equilibrium with coordinates $q=0$, $p=0$, then has the form
\begin{displaymath}
  H = \frac{1}{2} \langle Aw, w \rangle + \ldots
\end{displaymath}
where $A$ is the symmetric $2n \times 2n$-Hessian of $H$ at $0$ and the dots stand for terms of higher order in $w$. We now assume that the equilibrium point $w=0$ is elliptic, i.e. the spectrum of the linearized system, $\dot{w} = J A w$, is purely imaginary, spec$(J A) = \{ \pm i \l_1, \ldots, \pm i \l_n \}$ with real numbers $\l_1, \ldots, \l_n$. Here $J = \left( \begin{array}{cc} 0 & Id_n \\ -Id_n & 0 \end{array} \right)$ is the standard symplectic structure of $\R^{2n}$. If spec$(J A)$ is simple there exists a linear symplectic change of coordinates which brings the quadratic part of the Hamiltonian into normal form. Denoting the new coordinates by the same symbols as the old ones one has
\begin{displaymath}
  \langle Aw, w \rangle = \sum_{i=1}^n \l_i (q_i^2 + p_i^2).
\end{displaymath}

\begin{definition}
A Hamiltonian $H$ is in \emph{Birkhoff normal form up to order $m \geq 2$}, if it is of the form
\begin{equation} \label{hambirkhoffm}
  H = N_2 + N_4 + \ldots + N_m + H_{m+1} + \ldots,
\end{equation}
where the $N_k$, $2 \leq k \leq m$, are homogeneous polynomials of order $k$, which are actually functions of $q_1^2 + p_1^2$, \ldots, $q_n^2 + p_n^2$, and where $H_{m+1} + \ldots$ stands for arbitrary terms of order strictly greater than $m$. If this holds for any $m$, the Hamiltonian is simply said to be in \emph{Birkhoff normal form}.
\end{definition}

Note that if a Hamiltonian $H$ admits a Birkhoff normal form of order $m$, the coefficients of the expansion (\ref{hambirkhoffm}) up to order $m$ are uniquely determined, as long as the normalizing transformation is of the form id $+ \ldots$ \, . However, the normalizing transformation is by no means uniquely determined.

There are well known theorems guaranteeing the existence of a Birkhoff normal form up to order $m$ assuming that the frequencies $\l_1, \ldots, \l_n$ satisfy certain nonresonance conditions - see e.g. Theorem 4.3 in \cite{kapo}. We do not state these theorems here, because we will show the existence of a Birkhoff normal form up to order $4$ (in the case where $N$ is odd) of any periodic FPU chain by explicit calculations.

\section{Relative coordinates} \label{relcoord}

We start by expressing the FPU Hamiltonian $H_V$ in relative coordinates. Introduce $(v_1, \ldots, v_N) \in \R^N$ given by
\begin{displaymath}
  v_i := q_{i+1} - q_i \; (1 \leq i \leq N-1) \quad \textrm{and} \quad v_N := \frac{1}{N} \sum_{i=1}^N q_i.
\end{displaymath}
Then $v = M q$ is a linear change of the coordinates $q_1, \ldots, q_N$ with the $N \times N$-matrix
\begin{displaymath}
M = \left( \begin{array}{ccccc}
-1 & 1 & 0 & \ldots & 0 \\
0 & \ddots & \ddots &  & \vdots \\
\vdots &&&& 0 \\
0 & \ldots & 0 & -1 & 1 \\
N^{-1} & \ldots &  & \ldots & N^{-1} \\
\end{array} \right).
\end{displaymath}

The variables $u = (u_1, \ldots, u_N)$ conjugate to $v = (v_1, \ldots, v_N)$ are then given by $u = (M^T)^{-1} p$. $(M^T)^{-1}$ can be computed to be
\begin{equation} \label{minvformula}
(M^T)^{-1} = \frac{1}{N} \left( \begin{array}{ccccc}
1 & \ldots & \quad & \ldots & 1 \\
2 & \ldots &  & \ldots & 2 \\
\vdots & & & & \vdots \\
\vdots & & & & \vdots \\
N & \ldots &  & \ldots & N \\
\end{array} \right) - \left( \begin{array}{ccccc}
1 & 0 & \ldots & \ldots & 0 \\
1 & 1 & 0 & \ldots & 0 \\
\vdots & & & & \vdots \\
1 & \ldots & & 1 & 0 \\
0 & \ldots &  & \ldots & 0 \\
\end{array} \right).
\end{equation}

We already have mentioned that the total momentum $\sum_{j=1}^N p_j = u_N$ is conserved, and we write its constant value as $N \cdot P$. The motion of the center of mass $\frac{1}{N} \sum_{j=1}^N q_j = v_N$ is linear and therefore unbounded. It turns out that $H_V$ can be expressed as a function of the canonical variables $(v,u) = (v_k, u_k)_{\1N1}$ and $P$. To express $H_V$ in terms of $(v,u)$, note that by (\ref{minvformula}), $u_k = kP - \sum_{j=1}^k p_j$ for any $\1N1$. Hence
\begin{displaymath}
  p_1 = -u_1 + P; \, p_N = u_{N-1} + P; \, p_k = (u_{k-1} - u_k) + P \qquad (2 \leq k \leq N-1)
\end{displaymath}
and thus
\begin{displaymath}
\frac{1}{2} \sum_{j=1}^N p_j^2 = \frac{N P^2}{2} + \frac{1}{2} \left( u_1^2 + (u_1 - u_2)^2 + \ldots + (u_{N-2} - u_{N-1})^2 + u_{N-1}^2 \right).
\end{displaymath}
Moreover, using that $q_{N+1} - q_N = q_1 - q_N = -\sum_{k=1}^{N-1}(q_{k+1} - q_k)$ one gets for any $s \in \Z_{\geq 1}$
\begin{displaymath}
  \sum_{j=1}^N (q_{j+1} - q_j)^s = \sum_{k=1}^{N-1} v_k^s + \left( -\sum_{k=1}^{N-1} v_k \right)^s = \sum_{k=1}^{N-1} v_k^s + (-1)^s \left( \sum_{k=1}^{N-1} v_k \right)^s.
\end{displaymath}
Combining the two expressions displayed above yields
\begin{displaymath}
  H_V = \frac{NP^2}{2} + \tilde{H}_V,
\end{displaymath}
where
{\setlength\arraycolsep{1pt}\begin{eqnarray}
  \tilde{H}_V & = & \frac{1}{2} \left( u_1^2 \!+\! \sum_{l=1}^{N-2} (u_{l+1} \!-\! u_l)^2 \!+\! u_{N-1}^2 \right) + \frac{1}{2} \left( \sum_{k=1}^{N-1} v_k^2 + \left( \sum_{k=1}^{N-1} v_k \right)^2 \right) \label{hbavu}\\
&& \, + \frac{\a}{3!} \left( \sum_{k=1}^{N-1} v_k^3 \!-\! \left( \sum_{k=1}^{N-1} v_k \right)^3 \right) \!+\! \frac{\b}{4!} \left( \sum_{k=1}^{N-1} v_k^4 + \left( \sum_{k=1}^{N-1} v_k \right)^4 \right) + O(v^5). \nonumber
\end{eqnarray}}
Note that for any values of $P$, $\a$, and $\b$, the point $(v,u) = (0,0)$ is a critical point of the Hamiltonian $\hvt$. We will see in the next section that it is an elliptic fixed point.

\section{Linearized Birkhoff coordinates} \label{lbc}

We now compute the Birkhoff normal form of the FPU Hamiltonian $\hvt$ up to order $4$ near the fixed point $(v,u) = (0,0)$, taking the expansion (\ref{hbavu}) of $\hvt$ as a starting point. Write $\hvt$ as
\begin{equation} \label{hvdecomp}
\hvt = H_u + H_v,
\end{equation}
where $H_u$ and $H_v$ denote the $u$- and $v$-dependent parts of (\ref{hbavu}), respectively. Note that the Taylor expansion of $\hvt$ at $(v,u) = (0, 0)$ is \emph{not} in Birkhoff normal form up to order $2$. In a first step we therefore want to choose a linear canonical transformation $(\xi_k, \eta_k)_{\1N1} \mapsto (v_k, u_k)_{\1N1}$ so that when expressed in the new variables $(\xi, \eta) = (\xi_k, \eta_k)_{\1N1}$, the FPU Hamiltonian is in Birkhoff normal form up to order $2$.

The proposed canonical transformation has naturally come up in our earlier work on the Toda lattice \cite{ahtk3}. It turns out that (almost) the same transformation works for any FPU chain.

It is convenient to use complex notation for $\xi_k$, $\eta_k$ ($\1N1$),
\begin{displaymath}
  \z_k := \frac{1}{\sqrt{2}} (\xi_k - i \eta_k), \quad \z_{-k} := \frac{1}{\sqrt{2}} (\xi_k + i \eta_k).
\end{displaymath}
The minus sign in the definition of $\z_k$ is chosen so that $d\z_k \wedge d\z_{-k} = i d\xi_k \wedge d\eta_k$. The vector $\z = (\z_k)_{1 \leq |k| \leq N-1}$ is an element in the space
\begin{equation} \label{zetaspace}
  \mathcal{Z} := \{ z = (z_k)_{1 \leq |k| \leq N-1} \in \C^{2N-2} : z_{-k} = \overline{z_k} \quad \forall \; \1N1 \}.
\end{equation}
The components of $\z$ satisfy the identity
\begin{equation} \label{realcomplex}
  e^{i \pi j k/N} \z_k + e^{-i \pi j k/N} \z_{-k} = \sqrt{2} \, \left( \cos \left( \frac{j \pi k}{N} \right) \xi_k + \sin \left( \frac{j \pi k}{N} \right) \eta_k \right).
\end{equation}

For the rest of this paper, we use the notation
\begin{equation}
  \l_k := \big| \sin \frac{k\pi}{N} \big|^\frac{1}{2} \qquad (0 \leq |k| \leq N-1).
\end{equation}

The proposed transformation $\mathcal{Z} \to \R^{2N-2}$, $\z \mapsto (v,u)$ is given by the formulas
\setlength\arraycolsep{1pt}{\begin{eqnarray}
  u_1(\z) & = & \frac{1}{\sqrt{N}} \sum_{1 \leq |k| \leq N-1} \l_k \z_k, \label{transfu1} \\
u_{l+1}(\z) - u_l(\z) & = & \frac{1}{\sqrt{N}} \sum_{1 \leq |k| \leq N-1} \l_k e^{2\pi i lk/N} \z_k \quad (1 \! \leq \! l \! \leq \! N-2), \label{transfu2} \\
-u_{N-1}(\z) & = & \frac{1}{\sqrt{N}} \sum_{1 \leq |k| \leq N-1} \l_k e^{2\pi i (N-1) k/N} \z_k. \label{transfu3}
\end{eqnarray}}
and
\begin{equation} \label{vlzetak}
  v_l(\z) = \frac{1}{\sqrt{N}} \sum_{1 \leq |k| \leq N-1} \l_k e^{2\pi i lk/N} e^{-i\pi k/N} \z_k \quad (1 \leq l \leq N-1).
\end{equation}
Note that (\ref{transfu3}) is actually a consequence of (\ref{transfu1}) and (\ref{transfu2}): The left and right hand sides of (\ref{transfu1})-(\ref{transfu3}) both add up to $0$. From the fact that the Toda lattice is integrable it follows that the transformation above is a canonical linear isomorphism \cite{ahtk2}. To make this paper self-contained we directly verify that this is indeed the case.

\begin{lemma}
The linear transformation $\mathcal{Z} \to \R^{2N-2}$, $\z \mapsto (v,u)$, as defined by (\ref{transfu1})-(\ref{vlzetak}), is bijective and canonical.
\end{lemma}

\begin{proof}
First let us show
\begin{eqnarray}
\{ v_l(\z), u_m(\z) \} & = & i \, \delta_{lm}, \label{zzpm}\\
\{ v_l(\z), v_m(\z) & = & 0,  \label{zzpp}\\
\{ u_l(\z), u_m(\z) \} & = & 0 \label{zzmm}
\end{eqnarray}
for any $1 \leq l,m \leq N-1$. Since $(v,u)$ are canonical coordinates on $\R^{2(N-1)}$, the proof of (\ref{zzpm}) amounts to showing that
\begin{displaymath}
  \sum_{k=1}^{N-1} \left( \frac{\partial v_l}{\partial \z_k} \frac{\partial u_m}{\partial \z_{-k}} - \frac{\partial v_l}{\partial \z_{-k}} \frac{\partial u_m}{\partial \z_k} \right) = i \, \delta_{lm}
\end{displaymath}
for any $1 \leq l,m \leq N-1$. It follows from (\ref{transfu1})-(\ref{vlzetak}) that for any $\1N1$,
\begin{displaymath}
  \frac{\partial v_l}{\partial \z_k} = \frac{\l_k}{\sqrt{N}} e^{\pi i (2l-1) k/N}, \quad \frac{\partial v_l}{\partial \z_{-k}} = \frac{\l_k}{\sqrt{N}} e^{-\pi i (2l-1) k/N},
\end{displaymath}
\begin{displaymath}
  \frac{\partial u_m}{\partial \z_k} = \frac{\l_k}{\sqrt{N}} \sum_{j=0}^{m-1} e^{2 \pi i j k / N}, \quad \frac{\partial u_m}{\partial \z_{-k}} = \frac{\l_k}{\sqrt{N}} \sum_{j=0}^{m-1} e^{-2 \pi i j k / N}.
\end{displaymath}
Hence
\begin{eqnarray*}
  \frac{\partial v_l}{\partial \z_k} \frac{\partial u_m}{\partial \z_{-k}} & - & \frac{\partial v_l}{\partial \z_{-k}} \frac{\partial u_m}{\partial \z_k} \\
& = & \frac{\l_k^2}{N} \left( e^{\pi i (2l-1) k/N} \sum_{j=0}^{m-1} e^{-2 \pi i j k / N} - e^{-\pi i (2l-1) k/N} \sum_{j=0}^{m-1} e^{2 \pi i j k / N} \right) \\
& = & \frac{\l_k^2}{N} \sum_{j=0}^{m-1} \left( e^{\frac{\pi i k}{N} (2l-2j-1)} - e^{\frac{\pi i k}{N} (2j-2l+1)} \right) \\
& = & \frac{2i}{N} \sin \frac{k \pi}{N} \sum_{j=0}^{m-1} \sin \left( \frac{k \pi}{N} (2(l-j)-1) \right) \\
& = & \frac{i}{N} \sum_{j=0}^{m-1} \left( \cos \frac{2 k \pi (1 - (l-j))}{N} - \cos \frac{2 k \pi (l-j)}{N} \right),
\end{eqnarray*}
where for the latter identity we used that $2 \sin x \sin y = \cos(x-y) - \cos(x+ y)$. Taking the sum over $k$ and changing the order of summation then leads to
\begin{eqnarray*}
  \sum_{k=1}^{N-1} \left( \frac{\partial v_l}{\partial \z_k} \frac{\partial u_m}{\partial \z_{-k}} \!-\! \frac{\partial v_l}{\partial \z_{-k}} \frac{\partial u_m}{\partial \z_k} \right) & = & \frac{i}{N} \! \sum_{j=0}^{m-1} \! \sum_{k=1}^{N-1} \! \left( \cos \frac{2 k \pi (1 \!-\! (l\!\!-\!\!j))}{N} \!-\! \cos \frac{2 k \pi (l\!\!-\!\!j)}{N} \right) \\
& = & \frac{i}{N} \sum_{j=0}^{m-1} N (\delta_{l-j,1} - \delta_{l-j,0}) \\
& = & i \sum_{j=0}^{m-1} (\delta_{l,j+1} - \delta_{l,j}) = i (\delta_{lm} - \delta_{l0}) = i \delta_{lm},
\end{eqnarray*}
as claimed. To prove (\ref{zzpp}) and (\ref{zzmm}) one argues in a similar way. From (\ref{zzpm})- (\ref{zzmm}) it immediately follows that the linear map $\xi \mapsto (v,u)$ is bijective and canonical.
\end{proof}

We now compute $\hvt$ in terms of the new variables $\z$. According to the decomposition (\ref{hvdecomp}), we compute $H_u(\z)$ and $H_v(\z)$ separately. To obtain $H_u(\z)$, we substitute (\ref{transfu1})-(\ref{transfu3}) into the expression $\frac{1}{2} \left( u_1^2 + \sum_{l=1}^{N-2} (u_{l+1} - u_l)^2 + u_{N-1}^2 \right)$ and get
\begin{eqnarray*}
  H_u(\z) & = & \frac{1}{2N} \sum_{l=0}^{N-1} \left( \sum_{1 \leq |k| \leq N-1} \l_k e^{2\pi i lk/N} \z_k \right)^2 \\
& = & \frac{1}{2N} \sum_{1 \leq |k|, |k'| \leq N-1} \l_k \l_{k'} \left( \sum_{l=0}^{N-1} e^{2\pi i l(k+k')/N} \right) \z_k \z_{k'}.
\end{eqnarray*}
Using again that $\sum_{l=0}^{N-1} e^{2\pi i lk/N} = N \delta_{k0}$ and $\l_k = \l_{-k}$ for any $0 \leq |k| \leq N-1$, one obtains
\begin{displaymath}
  H_u(\z) = \sum_{k=1}^{N-1} \l_k^2 \z_k \z_{-k}.
\end{displaymath}

Before computing $H_v(\z)$, we simplify its expansion in terms of the variables $(v_k)_{\1N1}$. Define $v_0$ by the expression on the right hand side of (\ref{vlzetak}) evaluated at $l=0$. Note that
\begin{displaymath}
  \sum_{l=0}^{N-1} v_l = \frac{1}{\sqrt{N}} \sum_{1 \leq |k| \leq N-1} \l_k \z_k e^{-i\pi k/N} \left( \sum_{l=0}^{N-1} e^{2\pi i lk/N} \right) = 0.
\end{displaymath}
Hence $\sum_{l=1}^{N-1} v_l = -v_0$ and therefore
\begin{equation} \label{havkexp}
  H_v = \sum_{l=0}^{N-1} \left( \frac{1}{2} v_l^2 + \frac{\a}{3!} v_l^3 + \frac{\b}{4!} v_l^4 \right) + O(|v|^5).
\end{equation}
Substituting the expression (\ref{vlzetak}) for $v_l$ in the quadratic term in the expansion (\ref{havkexp}), we get
\begin{eqnarray*}
  \frac{1}{2} \sum_{l=0}^{N-1} v_l^2 & = & \frac{1}{2N} \sum_{1 \leq |k|, |k'| \leq N-1} \!\! \l_k \l_{k'} \left( \sum_{l=0}^{N-1} e^{2\pi i l(k+k')/N} \right) e^{-i\pi (k+k')/N} \z_k \z_{k'} \\
& = & \sum_{k=1}^{N-1} \l_k^2 \z_k \z_{-k},
\end{eqnarray*}
where we again used that $\l_k = \l_{-k}$ and $\sum_{l=0}^{N-1} e^{2\pi i lk/N} = N \delta_{k0}$ for any $0 \leq |k| \leq N-1$.

The terms of third and fourth order in $H_v$ are treated similarly. Combining the above computations leads to
\begin{displaymath}
  \hvt(\z) = G_2 + \a G_3 + \b G_4 + O(\z^5)
\end{displaymath}
with
\begin{eqnarray}
G_2 & := & 2 \sum_{k=1}^{N-1} \l_k^2 \z_k \z_{-k}, \label{g2lambdadef} \\
G_3 & := & -\frac{1}{6\sqrt{N}} \sum_{(k,k',k'') \in K_3} (-1)^{(k+k'+k'')/N} \l_k \l_{k'} \l_{k''} \z_k \z_{k'} \z_{k''}, \nonumber \\
G_4 & := & \frac{1}{24N} \!\! \sum_{(k,k',k'',k''') \in K_4} (-1)^{(k\!+\!k'\!+\!k''\!+\!k''')/N} \l_k \l_{k'} \l_{k''} \l_{k'''} \z_k \z_{k'} \z_{k''} \z_{k'''}, \label{g4lambdadef}
\end{eqnarray}
where
\begin{eqnarray}
  K_3 & := & \{ (k,k',k'') \in \Z^3: \, 1 \leq |k|, |k'|, |k''| \leq N-1 \label{indexk3def} \\
&& \qquad \qquad \qquad \qquad \qquad \textrm{ and } k+k'+k'' \equiv 0 \textrm{ mod } N \} \nonumber
\end{eqnarray}
and
\begin{eqnarray}
  K_4 & := & \{ (k,k',k'',k''') \in \Z^4: \, 1 \leq |k|, |k'|, |k''|, |k'''| \leq N-1 \label{indexk4def} \\
&& \qquad \qquad \qquad \qquad \qquad \textrm{ and } k+k'+k''+k''' \equiv 0 \textrm{ mod } N \}. \nonumber
\end{eqnarray}
Note that $G_2$, $G_3$, and $G_4$ are independent of $\a$ and $\b$. So indeed, $\hvt$ is in Birkhoff normal form up to order $2$, and it follows that $\z = 0$ is an elliptic fixed point of the corresponding Hamiltonian system.

\section{Proof of Theorem \ref{bnffputheorem}} \label{mainproof}

We now begin by transforming the $\hvt(\z)$ into its Birkhoff normal form up to order $4$. Here we follow a standard procedure - see e.g. section $14$ in \cite{kapo}. The phase space $\mathcal{Z}$, defined in (\ref{zetaspace}), is endowed with the Poisson bracket
\begin{displaymath}
  \{ F,G \} = i \sum_{1 \leq |k| \leq N-1} \sigma_k \frac{\partial F}{\partial \z_k} \frac{\partial G}{\partial \z_{-k}},
\end{displaymath}
where $\sigma_k = \textrm{sgn} \, (k)$ is the sign of $k$. The Hamiltonian vector field $X_F$ associated to the Hamiltonian $F$ is then given by $X_F = i \sum_{1 \leq |k| \leq N-1} \sigma_k \frac{\partial F}{\partial \z_{-k}} \frac{\partial}{\partial \z_k}$. With a first canonical transformation we want to eliminate the third order term $\a G_3$ in $\hvt(\z)$. By a by now standard precedure we construct such a canonical transformation on the phase space $\mathcal{Z}$ as the time-$1$-map $\Psi_1 := X^t_{\a F_3}|_{t=1}$ of the flow $X^t_{\a F_3}$ of a real analytic Hamiltonian $\a F_3$ which is a homogeneous polynomial in $\z_k$ ($1 \leq |k| \leq N-1$) of degree $3$ and solves the homological equation
\begin{equation} \label{homeqn}
  \{ G_2, \a F_3 \} + \a G_3 = 0.
\end{equation}
To simplify notation we momentarily write $F$ instead of $\a F_3$ and $H$ instead of $\hvt$. Assuming for the moment that (\ref{homeqn}) can be solved and that $X^t_F$ is defined for $0 \leq t \leq 1$ in some neighbourhood of the origin in $\mathcal{Z}$, we can use Taylor's formula to expand $H \circ X^t_F$ around $t=0$,
\begin{eqnarray}
  H \circ X^t_F & = & H \circ X^0_F + \int_0^t \frac{d}{ds}(H \circ X^s_F) ds \nonumber\\
& = & H + \int_0^t \{ H, F \} \circ X^s_F \, ds \nonumber\\
& = & H + t \, \{ H, F \} + \int_0^t ds \int_0^s ds' \frac{d}{ds'}(\{ H, F \} \circ X^{s'}_F) \nonumber\\
& = & H + t \, \{ H, F \} + \int_0^t (t-s) \{ \{ H, F \}, F \} \circ X^s_F \, ds. \label{hbataylort}
\end{eqnarray}
When evaluating this expression at $t=1$, one gets
\begin{eqnarray*}
  H \circ \Psi_1 & = & G_2 + \{ G_2, F \} + \int_0^1 (1-t) \{ \{ G_2, F \}, F \} \circ X^t_F dt \\
&& \quad + \a G_3 + \int_0^1 \{ \a G_3, F \} \circ X^t_F \, dt + \b G_4 + O(\z^5).
\end{eqnarray*}
Using that $\{ G_2, F \} + G_3 = 0$, the latter expression simplifies and we get
\begin{displaymath}
  H \circ \Psi_1 = G_2 + \int_0^1 t \, \{ \a G_3, F \} \circ X^t_F \, dt + \b G_4 + O(\z^5).
\end{displaymath}
Integrating by parts once more and taking into account that $F \equiv \a F_3$ is homogeneous of degree $3$ one obtains, in view of (\ref{hbataylort}),
\begin{equation} \label{H3}
  \hvt \circ \Psi_1 = G_2 + \frac{1}{2} \{ \a G_3, \a F_3 \} + \b G_4 + O(\z^5).
\end{equation}
Note that $\{ G_3, F_3 \}$ is homogeneous of order $4$. Hence our first step is achieved. It remains to solve (\ref{homeqn}). Since $G_3$ contains only monomials with $(k,k',k'') \in K_3$ (cf (\ref{indexk3def})), also $F_3$ need only contain such monomials,
\begin{displaymath}
  F_3 = \sum_{(k,k',k'') \in K_3} F^{(3)}_{kk'k''} \z_k \z_{k'} \z_{k''}
\end{displaymath}
which leads to
\begin{eqnarray}
\{ G_2, F_3 \} & = & i \sum_{1 \leq |k| \leq N-1} 2 \sigma_k \l_k^2 \z_{-k} \frac{\partial F_3}{\partial \z_{-k}} \nonumber\\
& = & -i \sum_{(k,k',k'') \in K_3} (s_k + s_{k'} + s_{k''}) F^{(3)}_{kk'k''} \z_k \z_{k'} \z_{k''}, \label{g2f3poisson}
\end{eqnarray}
where
\begin{displaymath}
s_k := 2 \sigma_k \l_k^2 = 2 \sin \frac{k \pi}{N}.
\end{displaymath}

The following result is due to Beukers and Rink (cf. \cite{rink01, rink06}):
\begin{lemma} \label{nonres3lemma}
(\cite{rink01, rink06}) For any $(k,k',k'') \in K_3$,
\begin{displaymath}
s_k + s_{k'} + s_{k''} \neq 0.
\end{displaymath}
\end{lemma}

Let us remark that Lemma \ref{nonres3lemma} also follows from the integrability of the Toda lattice (cf. \cite{ahtk3}). We include the self-contained proof due to Beukers and Rink.

\begin{proof}
Suppose that $(k,k',k'') \in K_3$ satisfies $s_k + s_{k'} + s_{k''} = 0$. It follows from $k+k'+k'' \equiv 0$ mod $N$ that either $s_{k''} = -s_{k+k'}$ or $s_{k''} = s_{k+k'}$, according to whether $k+k'+k'' \equiv 0$ or $k+k'+k'' \equiv N$ mod $2N$.

In the first case, it follows that
\begin{equation} \label{resrelcase1}
2i\sin\frac{k\pi}{N} + 2i\sin\frac{k'\pi}{N} - 2i\sin\left( \frac{k\pi}{N} + \frac{k'\pi}{N} \right) = 0.
\end{equation}
Setting $x := e^\frac{ik\pi}{N}$ and $y := e^\frac{ik'\pi}{N}$, one can rewrite (\ref{resrelcase1}) as
\begin{equation} \label{beukerstrickeqn}
  0 = x - \frac{1}{x} + y - \frac{1}{y} - xy + \frac{1}{xy} = (1-x) (1-y) (1-xy)\frac{1}{xy}.
\end{equation}
It follows that any solution of (\ref{beukerstrickeqn}) contradicts the assumption $1 \leq |k|, |k'|, |k''| \leq N-1$. Indeed, solutions with $x=1$ (i.e. $k \equiv 0$ mod $2N$), $y=1$ (i.e. $k' \equiv 0$ mod $2N$), or $xy=1$ (i.e. $k+k' \equiv 0$ mod $2N$ and thus $k'' \equiv 0$ mod $2N$), contradict this assumption.

In the second case, we have instead of (\ref{resrelcase1})
\begin{equation} \label{resrelcase2}
2i\sin\frac{k\pi}{N} + 2i\sin\frac{k'\pi}{N} + 2i\sin\left( \frac{k\pi}{N} + \frac{k'\pi}{N} \right) = 0.
\end{equation}
With $x,y$ as above, it now follows from (\ref{resrelcase2}) that
\begin{displaymath}
  0 = x - \frac{1}{x} + y - \frac{1}{y} + xy - \frac{1}{xy} = -(1+x) (1+y) (1-xy)\frac{1}{xy}.
\end{displaymath}
Again we conclude that any solution of (\ref{resrelcase2}) contradicts the assumption  $1 \leq |k|, |k'|, |k''| \leq N-1$. Indeed, solutions with $x=-1$ (i.e. $k \equiv N$ mod $2N$), $y=-1$ (i.e. $k' \equiv N$ mod $2N$), or $xy=1$ (i.e. $k+k' \equiv 0$ mod $2N$ and thus $k'' \equiv N$ mod $2N$), contradict this assumption.
\end{proof}

By Lemma \ref{nonres3lemma}, one can define $F_3$ as follows
\begin{displaymath}
  i F^{(3)}_{kk'k''} := \left\{  \begin{array}{ll}
\frac{G^{(3)}_{kk'k''}}{s_k + s_{k'} + s_{k''}} & \qquad (k,k',k'') \in K_3 \\
& \\
0 & \qquad \textrm{otherwise}
\end{array} \right.
\end{displaymath}
Then $\{ G_2, \a F_3 \} + \a G_3 = 0$. Written more explicitly, the nonzero coefficients of $F_3$ are
\begin{displaymath}
  i F^{(3)}_{kk'k''} = -\frac{(-1)^{(k+k'+k'')/N}}{6 \sqrt{N}} \frac{\sqrt{|\sk \skk \skkk|}}{2\sk + 2\skk + 2\skkk}.
\end{displaymath}

In a second step we normalize the $4$th order term $\b G_4 + \frac{\a^2}{2} \{ G_3, F_3 \}$ in (\ref{H3}). We decompose this sum into its contibution to the Birkhoff normal form and the rest, to be transformed away in a moment. Let us first compute $\frac{\a^2}{2} \{ G_3, F_3 \}$ in a more explicit form:

\begin{eqnarray*}
  \frac{\a^2}{2} \{ G_3, F_3 \} & = & \frac{i}{2} \, \a^2 \!\!\!\! \sum_{1 \leq |k| \leq N-1} \!\! \sigma_k \frac{\partial G_3}{\partial \z_k} \frac{\partial F_3}{\partial \z_{-k}} = \frac{1}{2} \, \a^2 \!\!\!\! \sum_{1 \leq |k| \leq N-1} \!\! \sigma_k \frac{\partial G_3}{\partial \z_k} \frac{\partial (i F_3)}{\partial \z_{-k}} \\
& = & \frac{\a^2}{2N} \sum_{1 \leq |k| \leq N-1} \sigma_k \left( \frac{3}{6} \!\!\!\!\!\! \sum_{1 \leq |l|, |m| \leq N-1, \atop l+m = -k+rN} \!\!\!\! \, (-1)^r \l_k \l_l \l_m \z_l \z_m \right) \\
&& \qquad \qquad \cdot \left( \frac{3}{6} \!\!\!\!\!\! \sum_{1 \leq |l'|, |m'| \leq N-1, \atop l'+m = k+r'N} \!\!\!\! (-1)^{r'} \frac{\l_k \l_{l'} \l_{m'}}{s_{-k} \!+\! s_{l'} \!+\! s_{m'}} \z_{l'} \z_{m'} \right) \\
& = & \frac{\a^2}{16N} \!\!\!\! \sum_{1 \leq |k| \leq N-1} \sum_{1 \leq |l|, |m|, |l'|, |m'| \leq N-1 \atop {l+m-rN = -k \atop l'+m'-r'N = k}} \!\!\!\!\!\!\!\! (-1)^{r+r'} \frac{s_k \l_l \l_m \l_{l'} \l_{m'}}{s_{-k} + s_{l'} + s_{m'}} \z_l \z_m \z_{l'} \z_{m'},
\end{eqnarray*}
where for the latter equality we used that $2 \sigma_k \l_k^2 = s_k$. Setting
\begin{equation} \label{elmlmdef}
\e_{lml'm'} := \frac{l+m+l'+m'}{N}
\end{equation}
one then gets
\begin{eqnarray*}
  \frac{\a^2}{2} \{ G_3, F_3 \} & = & \frac{\a^2}{16N} \!\!\!\! \sum_{1 \leq |k| \leq N-1} \sum_{l+m \equiv -k \, mod \, N \atop l'+m'\equiv k \, mod \, N} \!\!\!\! (-1)^{\e_{lml'm'}} \frac{\l_l \l_m \l_{l'} \l_{m'}}{-1 \!+\! (s_{l'} \!\!+\!\! s_{m'})/s_k} \z_l \z_m \z_{l'} \z_{m'} \\
& = & \frac{\a^2}{16N} \sum_{k=1}^{N-1} \sum_{l+m \equiv -k \, mod \, N \atop l'+m' \equiv k \, mod \, N} \!\!\!\! (-1)^{\e_{lml'm'}} \frac{\l_l \l_m \l_{l'} \l_{m'}}{-1 + (s_{l'} + s_{m'})/s_k} \z_l \z_m \z_{l'} \z_{m'} \\
&& + \frac{\a^2}{16N} \sum_{k=1}^{N-1} \sum_{l+m \equiv k \, mod \, N \atop l'+m'\equiv -k \, mod \, N} \!\!\!\! (-1)^{\e_{lml'm'}} \frac{\l_l \l_m \l_{l'} \l_{m'}}{-1 - (s_{l'} + s_{m'})/s_k} \z_l \z_m \z_{l'} \z_{m'} \\
& = & \frac{\a^2}{16N} \sum_{k=1}^{N-1} \sum_{l+m \equiv -k \, mod \, N \atop l'+m' \equiv k \, mod \, N} \!\!\! \left( \frac{1}{-1 + (s_{l'} \!\!+\!\! s_{m'})/s_k} + \frac{1}{-1 - (s_l \!\!+\!\! s_m)/s_k} \right) \\ 
&& \qquad \qquad \qquad \qquad \cdot \, (-1)^{\e_{lml'm'}} \l_l \l_m \l_{l'} \l_{m'} \z_l \z_m \z_{l'} \z_{m'}.
\end{eqnarray*}
Note that for $k = l'+m'+r'N$ with $\1N1$ and $r' \in \Z$ we have
\begin{displaymath}
  s_k = |s_{l'+m'}|.
\end{displaymath}
Introduce\footnote{To keep the formula for $c_{lml'm'}$ as simple as possible we have not symmetrized the coefficients $c_{lml'm'}$.} for any $(l,m,l',m') \in K_4$
\begin{equation} \label{cklmndef}
  c_{lml'm'} = \left\{ \begin{array}{ll}
 \frac{1}{-1 + \frac{s_{l'} + s_{m'}}{|s_{l'+m'}|}} - \frac{1}{1 + \frac{s_l + s_m}{|s_{l+m}|}} & \quad \textrm{if } l+m \not \equiv 0 \; \textrm{mod} \, N \\
&\\
0 & \quad \textrm{otherwise.} \end{array} \right.
\end{equation}
We then get
\begin{equation} \label{g3f3}
  \frac{\a^2}{2} \{ G_3, F_3 \} = \frac{\a^2}{16N} \sum_{ \atop (l,m,l',m') \in K_4} c_{lml'm'} (-1)^{\e_{lml'm'}} \l_l \l_m \l_{l'} \l_{m'} \z_l \z_m \z_{l'} \z_{m'}.
\end{equation}
Combined with formula (\ref{g4lambdadef}) for $G_4$, the quantity $\b G_4 + \frac{\a^2}{2} \{ G_3, F_3 \}$ becomes
\begin{equation} \label{4thordercomplete}
\frac{1}{24N} \!\!\!\! \sum_{ \atop (k,k',k'',k''') \in K_4} \!\!\!\! (-1)^{\e_{kk'k''k'''}} (\b \!+\! \frac{3\a^2}{2} c_{kk'k''k'''}) \l_k \l_{k'} \l_{k''} \l_{k'''} \z_k \z_{k'} \z_{k''} \z_{k'''}.
\end{equation}
We now decompose (\ref{4thordercomplete}) into its contribution to the Birkhoff normal form of $H_V$ and the rest, and we denote by $\pi_N$ the projection onto the former one, whereas the latter one will be (partially) transformed away by a second transformation $\Psi_2$.

\begin{lemma}
The normal form part of $\b G_4 + \frac{\a^2}{2} \{ G_3, F_3 \}$ is given by
\begin{eqnarray}
\pi_N \Big( \b G_4 & + & \frac{\a^2}{2} \{ G_3, F_3 \} \Big) \nonumber\\
& = & \! \frac{1}{4N} \! \left( \! \sum_{l=1}^{N-1} (\a^2 \!+\! (\b \!\!-\!\! \a^2) \l_l^4) |\z_l|^4 \!+\! 2 \!\!\!\!\!\!\!\!\!\! \sum_{1 \leq l \neq m \leq N-1} \!\!\!\!\!\!\!\! (\b \!\!-\!\! \a^2) \l_l^2 \l_m^2 |\z_l|^2 |\z_m|^2 \! \right) \!\! . \label{ping4ping3f3}
\end{eqnarray}
\end{lemma}

\begin{proof}
The indices $k,k',k'',k'''$ of the terms in $\b G_4 + \frac{\a^2}{2} \{ G_3, F_3 \}$ contributing to the normal form satisfy $(k,k',k'',k''') \in K_4^N$, where
\begin{eqnarray}
K_4^N & := & \{ (k,k',k'',k''') \in K_4 | \; \exists \, 1 \leq l \leq m \leq N-1 \textrm{ such that } \nonumber\\
&& \qquad \qquad \qquad \qquad \qquad \{ k,k',k'',k''' \} = \{ l,-l,m,-m \} \}. \label{nfcond}
\end{eqnarray}
In the case $l=m$, $\{ l,-l,l,-l \}$ in (\ref{nfcond}) is viewed as a set-like object whose two elements $l$ and $-l$ each have multiplicity two.

We investigate $\pi_N(G_4)$ and $\pi_N(\frac{1}{2} \{ G_3, F_3 \})$ separately. Let us start with $G_4$. We distinguish the cases $l=m$ and $l \neq m$ in $K_4^N$. For $l=m$, there are ${4 \choose 2} = 6$ distinct permutations of $(k,k',k'',k''')$ in $K_4^N$, whereas for $l \neq m$, all $4! = 24$ permutations of $(l,m,-l,-m)$ are distinct. Hence we have
\begin{eqnarray}
  \pi_N(\b G_4) & = & \frac{\b}{24N} \left( 6 \sum_{l=1}^{N-1} \l_l^4 |\z_l|^4 + 24 \!\!\!\! \sum_{1 \leq l < m \leq N-1} \!\!\!\! \l_l^2 \l_m^2 |\z_l|^2 |\z_m|^2 \right) \nonumber\\
& = & \frac{\b}{4N} \left( \sum_{l=1}^{N-1} \l_l^4 |\z_l|^4 + 2 \!\!\!\! \sum_{1 \leq l \neq m \leq N-1} \!\!\!\! \l_l^2 \l_m^2 |\z_l|^2 |\z_m|^2 \right). \label{ping4formula}
\end{eqnarray}

Now let us compute $\pi_N(\frac{1}{2} \{ G_3, F_3 \})$. We have to single out the matches of (\ref{nfcond}) for which in addition the coefficient $c_{kk'k''k'''}$ in (\ref{g3f3}) does not vanish, i.e.
\begin{displaymath}
k+k' \not \equiv 0 \, \textrm{mod} \; N \; \textrm{and} \; \; k+k'+k''+k''' \equiv 0 \; \textrm{mod} \; N.
\end{displaymath}
Hence there are two quadruples $(k,k',k'',k''')$ in $K_4^N$ which satisfy these additional conditions,
\begin{equation} \label{kmlnknlm}
\left. \begin{array}{rl}
k+k'' & = 0 \\
k'+k''' & = 0 \end{array} \right. \qquad \qquad \textrm{or} \qquad \qquad \left. \begin{array}{rl}
k+k''' & = 0 \\
k'+k'' & = 0 \end{array} \right..
\end{equation}
In both cases, we have $s_{k''} + s_{k'''} = -(s_k + s_{k'})$, and therefore (\ref{cklmndef}) reduces to
\begin{equation} \label{cklmnnfterms}
  c_{kk'k''k'''} = \frac{-2 |s_{k+k'}|}{|s_{k+k'}| + s_k + s_{k'}}.
\end{equation}
Note that (\ref{cklmnnfterms}) remains valid for $k+k' = N$, since in this case $s_{k+k'} = 0$ and $s_k + s_{k'} > 0$ as $k$ and $k'$ must satisfy $1 \leq k,k' \leq N-1$, but not for $k+k' = 0$, since in this case $|s_{k+k'}| + s_k + s_{k'} = 0$.

We first compute the diagonal part of $\pi_N \big( \frac{1}{2} \{ G_3, F_3 \} \big)$. In this case, the two possibilities in (\ref{kmlnknlm}) coincide and the solutions are
\begin{equation} \label{diaggf3coeff}
(k,k',k'',k''') = \left\{ \begin{array}{cccc} 
  (l, & l, & -l, & -l) \\
  (-l, & -l, & l, & l) \end{array} \right.,
\end{equation}
where $1 \leq l \leq N-1$.
The sum of the coefficients $c_{kk'k''k'''}$ for the two cases listed in (\ref{diaggf3coeff}) is
\begin{displaymath}
  c_{l,l,-l,-l} + c_{-l,-l,l,l} = -2 |s_{2l}| \left( \frac{1}{|s_{2l}| \!+\! 2 s_l} + \frac{1}{|s_{2l}| \!-\! 2 s_l} \right) = \frac{-4 s_{2l}^2}{s_{2l}^2 \!-\! 4 s_l^2} = 4 \cot^2 \frac{l \pi}{N}.
\end{displaymath}

We now turn to the off-diagonal part of $\pi_N \big( \frac{1}{2} \{ G_3, F_3 \} \big)$. The quadruples $(k,k',k'',k''') \in K_4$ satisfying (\ref{kmlnknlm}) for given $\{ l, m \} \subseteq \{ 1, \ldots, N-1 \}$ with $l < m$, $(k,k') = (\pm l, \pm m)$, and $(k'',k''') = (\pm l, \pm m)$, are
\begin{equation} \label{offdiaggf3coeff}
(k,k',k'',k''') = \left\{ \begin{array}{cccc} 
  (l, & m, & -l, & -m) \\
  (l, & -m, & -l, & m) \\
  (-l, & m, & l, & -m) \\
  (-l, & -m, & l, & m) \end{array} \right..
\end{equation}
The remaining matches are obtained from (\ref{offdiaggf3coeff}) by permuting the first and second or the third and fourth columns on the right hand side of (\ref{offdiaggf3coeff}), bringing the total number of all matches to $16 = 4 \cdot 4$. Note that by formula (\ref{cklmnnfterms}), these permutations leave the value of the coefficients $c_{kk'k''k'''}$ invariant. Taking the sum of the coefficients $c_{kk'k''k'''}$ for all the quadruples listed in (\ref{offdiaggf3coeff}), we obtain
\begin{eqnarray*}
 4 (c_{l,m,-l,-m} & + & c_{l,-m,-l,m} + c_{-l,m,l,-m} + c_{-l,-m,l,m}) \\
& = & -8 \Bigg( \frac{|s_{l+m}|}{|s_{l+m}| + s_l + s_m} + \frac{|s_{l-m}|}{|s_{l-m}| + s_l - s_m} \\ && \quad + \frac{|s_{l-m}|}{|s_{l-m}| - s_l + s_m} +
 \frac{|s_{l+m}|}{|s_{l+m}| - s_l - s_m} \Bigg) \\
& = & -16 \left( \frac{s_{l-m}^2}{s_{l-m}^2 - (s_l - s_m)^2} + \frac{s_{l+m}^2}{s_{l+m}^2 - (s_l + s_m)^2} \right) \\
& = & \frac{-16 (2 s_{l-m}^2 s_{l+m}^2 - s_{l-m}^2 (s_l + s_m)^2 - s_{l+m}^2 (s_l - s_m)^2)}{s_{l \!-\! m}^2 s_{l \!+\! m}^2 + (s_l \!-\! s_m)^2 (s_l \!+\! s_m)^2 \!-\! s_{l \!-\! m}^2 (s_l \!+\! s_m)^2 \!-\! s_{l \!+\! m}^2 (s_l \!-\! s_m)^2} \\
& = & -16,
\end{eqnarray*}
since $s_{l-m}^2 s_{l+m}^2 = (s_l-s_m)^2 (s_l+s_m)^2$. Collecting terms, we thus have
\begin{eqnarray}
  \pi_N \left( \frac{\a^2}{2} \{ G_3, F_3 \} \right) & = & \frac{\a^2}{16N} \left( \sum_{l=1}^{N-1} 4 \cos^2 \frac{\pi l}{N} |\z_l|^4 - 16 \!\!\!\!\!\! \sum_{1 \leq l < m \leq N-1} \!\!\!\!\!\! \l_l^2 \l_m^2 |\z_l|^2 |\z_m|^2 \right) \nonumber\\
& = & \frac{\a^2}{4N} \left( \sum_{l=1}^{N-1} (1 - \l_l^4) |\z_l|^4 - 2 \!\!\!\!\!\! \sum_{1 \leq l \neq m \leq N-1} \!\!\!\!\!\! \l_l^2 \l_m^2 |\z_l|^2 |\z_m|^2 \right). \label{ping3f3formula}
\end{eqnarray}
Adding up (\ref{ping4formula}) and (\ref{ping3f3formula}), we obtain (\ref{ping4ping3f3}).
\end{proof}

Now we want to remove [as much as possible of] the term $(\textrm{Id} - \pi_N) (\b G_4 + \frac{\a^2}{2} \{ G_3, F_3 \})$ from the Hamiltonian $\hvt \circ \Psi_1$ given by (\ref{H3}) by a second coordinate transformation $\Psi_2$. In view of formulas (\ref{g4lambdadef}) and (\ref{g3f3}) for $G_4$ and $\frac{1}{2} \{ G_3, F_3 \}$, respectively, and in complete analogy to the first step we look for a transformation $\Psi_2$ of the form $\Psi_2 = X_{F_4}^t|_{t=1}$ with
\begin{displaymath}
  F_4 = \sum_{(k,k',k'',k''') \in K_4 \setminus K_4^N} F^{(4)}_{kk'k''k'''} \z_k \z_{k'} \z_{k''} \z_{k'''},
\end{displaymath}
where $F^{(4)}_{\sigma(k,k',k'',k''')} = F^{(4)}_{(k,k',k'',k''')}$ for any permutation $\sigma(k,k',k'',k''')$ of the quadruple $(k,k',k'',k''') \in K_4 \setminus K_4^N$. We would like to determine the coefficients of $F_4$ in such a way that
\begin{equation} \label{g2f4eqn}
  \{ G_2, F_4 \} = -(\textrm{Id}-\pi_N) (\b G_4 + \frac{\a^2}{2} \{ G_3, F_3 \}).
\end{equation}
As in (\ref{g2f3poisson}) one gets
\begin{equation}
  \{ G_2, F_4 \} = -i \!\! \sum_{(k,k',k'',k''') \in K_4 \setminus K_4^N} \!\! (s_k + s_{k'} + s_{k''} + s_{k'''}) F^{(4)}_{kk'k''k'''} \z_k \z_{k'} \z_{k''} \z_{k'''}, \label{g2f4poisson}
\end{equation}
and equation (\ref{g2f4eqn}) combined with (\ref{4thordercomplete}) leads to
\setlength\arraycolsep{1pt}{\begin{eqnarray}
i (s_k & + &  s_{k'} + s_{k''} + s_{k'''}) F^{(4)}_{kk'k''k'''} \label{fkkkkeqn} \\
& = & \frac{1}{24N} (-1)^{\e_{kk'k''k'''}} (\b + \frac{3\a^2}{2} c_{kk'k''k'''}^S) \cdot \l_k \l_{k'} \l_{k''} \l_{k'''} \nonumber
\end{eqnarray}}
for any quadruple $(k,k',k'',k''')$ in $K_4 \setminus K_4^N$. Here $c_{kk'k''k'''}^S$ denotes the symmetrized version of $c_{kk'k''k'''}$,
\begin{equation} \label{symmcoeff}
  c_{kk'k''k'''}^S := \frac{1}{4!} \sum_{\sigma \in S_4} c_{\sigma(k,k',k'',k''')}.
\end{equation}
The following lemma due to Beukers and Rink (cf \cite{rink01}) determines the quadruples $(k,k',k'',k''') \in K_4 \setminus K_4^N$ for which $s_k + s_{k'} + s_{k''} + s_{k'''} = 0$. Let us introduce
\begin{displaymath}
  K_4^{res} := K_{res}^+ \cup K_{res}^- \subseteq K_4
\end{displaymath}
where
\setlength\arraycolsep{2pt}{
\begin{eqnarray*} 
   K_{res}^\pm & := & \Big\{ (k,k',k'',k''') \in K_4 | \; \exists \, l \in \N: 1 \leq l \leq \frac{N}{4} \;\; \textrm{so that} \\
&& \quad \{ k,k',k'',k''' \} = \{ \pm l, \pm l \mp N, \frac{N}{2} \mp l,-\frac{N}{2} \mp l \} \Big\}.
\end{eqnarray*}}
Note that if $N$ is odd, then $K_{res} = \emptyset$.

\begin{lemma} \label{nonres4lemma}
(\cite{rink01}) Let $(k_1,k_2,k_3,k_4) \in K_4 \setminus K_4^N$. Then
\begin{displaymath}
 s_k + s_{k'} + s_{k''} + s_{k'''} = 0 \textrm{ if and only if } (k,k',k'',k''') \in K_4^{res}.
\end{displaymath}
In particular, if $N$ is odd, then $s_k + s_{k'} + s_{k''} + s_{k'''} \neq 0$.
\end{lemma}
For the convenience of the reader a detailed proof of Lemma \ref{nonres4lemma} is given in Appendix \ref{nonresapp}.

By Lemma \ref{nonres4lemma}, if $N$ is odd, (\ref{fkkkkeqn}) can be solved for any $(k,k',k'',k''') \in K_4 \setminus K_4^N$ determining the coefficients $F_{(k,k',k'',k''')}$ with $(k,k',k'',k''') \in K_4 \setminus K_4^N$ in such a way that $F^{(4)}_{\sigma(k,k',k'',k''')} = F^{(4)}_{(k,k',k'',k''')}$ for any permutation $\sigma(k,k',k'',k''')$ of $(k,k',k'',k''') \in K_4 \setminus K_4^N$. With this choice of $F_4$ the canonical transformation $\Psi_2$ is then defined by $X_{F_4}^t|_{t=1}$. Composing $\Psi_1$ and $\Psi_2$, we obtain the transformation $\Xi := \Psi_1 \circ \Psi_2$. We have proved the following

\begin{prop} \label{localprop}
Assume that $N \geq 3$ is odd. The real analytic symplectic coordinate transformation $\z = \Xi(z)$, defined in a neighborhood of the origin in $\mathcal{Z}$, transforms the Hamiltonian $\hvt$ into its Birkhoff normal form up to order $4$. More precisely,
\begin{equation} \label{hamtaylorwk}
  \hvt \circ \Xi = G_2 + \pi_N \left( \b G_4 + \frac{\a^2}{2} \{ G_3, F_3 \} \right) + O(z^5),
\end{equation}
with $G_2$ and $\pi_N(\b G_4 + \frac{\a^2}{2} \{ G_3, F_3 \})$ given by (\ref{g2lambdadef}) and (\ref{ping4ping3f3}), respectively.
\end{prop}

Theorem \ref{bnffputheorem} can now be proved easily.
\begin{proof}[Proof of Theorem \ref{bnffputheorem}]
Proposition \ref{localprop} provides the Taylor series expansion of $\hvt$ in terms of the actions
\begin{equation} \label{ikdef}
  I = (I_k)_{\1N1}, \qquad I_k = \frac{x_k^2 + y_k^2}{2}.
\end{equation}
More precisely, $\hvt \circ \Xi = \Hab(I) + O(z^5)$, where $\Hab(I)$ is defined by
\begin{equation} \label{bnffpuodd}
  2 \sum_{k=1}^{N-1} \l_k^2 \, I_k + \frac{1}{4N} \sum_{k=1}^{N-1} (\a^2 + (\b - \a^2) \l_k^4) I_k^2 + \frac{\b - \a^2}{2N} \!\!\!\! \sum_{l \neq m \atop 1 \leq l,m \leq N-1} \!\!\!\! \l_l^2 \l_m^2 I_l I_m
\end{equation}
and $\l_k = |\sin \frac{k \pi}{N}|^\frac{1}{2}$. This proves Theorem \ref{bnffputheorem}.
\end{proof}

\section{Proof of Theorem \ref{bnffputheoremeven}} \label{evenproof}

Now we assume that $N$ is even. To obtain the normal form of the FPU Hamiltonian as claimed in Theorem \ref{bnffputheoremeven} we continue the investigations of the previous section. According to Lemma \ref{nonres4lemma}, equation (\ref{fkkkkeqn}) might have no solution $F^4_{kk'k''k'''}$ for $(k,k',k'',k''') \in K_4^{res}$. We first compute the projection $\pi_{res} (\b G_4 + \frac{\a^2}{2} \{ G_3, F_3 \})$ of $\b G_4 + \frac{\a^2}{2} \{ G_3, F_3 \}$ onto those terms which are indexed by quadruples $(k,k',k'',k''') \in K_4^{res}$, i.e. the projection onto the resonant non-normal form part of $\b G_4 + \frac{\a^2}{2} \{ G_3, F_3 \}$.

\begin{lemma} \label{resprojlemma}
Assume that $N$ is even. The resonant non-normal form part of $\b G_4 + \frac{\a^2}{2} \{ G_3, F_3 \}$ is given by
\begin{equation} \label{reszetaformula}
  \pi_{res} \left( \b G_4 + \frac{\a^2}{2} \{ G_3, F_3 \} \right) = -\frac{\b - \a^2}{4N} (R + R_\frac{N}{4})
\end{equation}
where
\begin{equation} \label{rresdef}
R := \sum_{1 \leq l < \frac{N}{4}} \!\! s_{2l} \left( \z_l \z_{-N+l} \z_{\frac{N}{2}-l} \z_{-\frac{N}{2}-l} + \z_{-l} \z_{N-l} \z_{\frac{N}{2}+l} \z_{-\frac{N}{2}+l} \right)
\end{equation}
with $c_k := 2 \cos \frac{k \pi}{N}$ for $\1N1$, and
\begin{equation} \label{rn4def}
R_{\frac{N}{4}} = \left \{ \begin{array}{ll}
\frac{1}{2} \left( \z_{\frac{N}{4}}^2 \z_{-\frac{3N}{4}}^2 + \z_{\frac{3N}{4}}^2 \z_{-\frac{N}{4}}^2 \right) & \quad \textrm{if } \frac{N}{4} \in \N \\
&\\
0 & \quad \textrm{otherwise.} \end{array} \right.
\end{equation}
\end{lemma}

\begin{proof}
Consider the formula (\ref{4thordercomplete}) for $\b G_4 + \frac{\a^2}{2} \{ G_3, F_3 \}$. At this point we need to consider the symmetrized version (\ref{symmcoeff}) of the coefficients $c_{klk'l'}$ defined by (\ref{cklmndef}). We claim that for any $(k_1,k_2,k_3,k_4) \in K_4^{res}$
\begin{equation} \label{cklmnressum}
  c_{k_1 k_2 k_3 k_4}^S = \frac{1}{4!} \sum_{\sigma \in S_4} c_{k_{\sigma(1)} k_{\sigma(2)} k_{\sigma(3)} k_{\sigma(4)}} = -\frac{2}{3}.
\end{equation}
Observe that $c_{k_1 k_2 k_3 k_4}$ is invariant under the transpositions $k_1 \leftrightarrow k_2$ and $k_3 \leftrightarrow k_4$. Hence (\ref{cklmnressum}) follows once we prove that
\begin{equation} \label{cklmnressum2}
  4 \left( c_{k_1 k_2 k_3 k_4} \!+\! c_{k_1 k_3 k_2 k_4} \!+\! c_{k_1 k_4 k_2 k_3} \!+\! c_{k_2 k_4 k_1 k_3} \!+\! c_{k_2 k_3 k_1 k_4} \!+\! c_{k_3 k_4 k_1 k_2} \right) \!=\! -16.
\end{equation}
Note that any element $(k_1, k_2, k_3, k_4) \in K_4^{res}$ is, mod $2N$, a permutation of an element of the form $(l,-N+l, N/2+l,-N/2+l)$ with $1 \leq |l| \leq N/4$. For such quadruples one gets by a straightforward computation
\begin{displaymath}
  c_{k_1 k_2 k_3 k_4} + c_{k_3 k_4 k_1 k_2} = -2-2 = -4
\end{displaymath}
and, with $c_l= 2 \cos \frac{l \pi}{N}$,
\begin{displaymath}
  c_{k_1 k_3 k_2 k_4} + c_{k_2 k_4 k_1 k_3} = - \frac{4}{2 + (s_l + c_l)} - \frac{4}{2 - (s_l + c_l)} = - \frac{8}{s_{2l}}
\end{displaymath}
as well as
\begin{displaymath}
  c_{k_1 k_4 k_2 k_3} + c_{k_2 k_3 k_1 k_4} = - \frac{4}{2 + (s_l - c_l)} - \frac{4}{2 - (s_l - c_l)} = \frac{8}{s_{2l}}.
\end{displaymath}
Substituting these three identities into the left hand side of (\ref{cklmnressum2}) leads to the claimed identity (\ref{cklmnressum2}).

Moreover, by the definition (\ref{elmlmdef}) of $\e_{lml'm'}$ one has for any $(k_1,k_2,k_3,k_4) \in K_4^{res}$ and any $\sigma \in S_4$, that $\e_{k_{\sigma(1)} k_{\sigma(2)} k_{\sigma(3)} k_{\sigma(4)}} = \pm 1$ and hence
\begin{displaymath}
(-1)^{\e_{k_{\sigma(1)} k_{\sigma(2)} k_{\sigma(3)} k_{\sigma(4)}}} = -1.
\end{displaymath}
Further,
\begin{displaymath}
\l_{k_1} \l_{k_2} \l_{k_3} \l_{k_4} = \left| \sin \frac{l \pi}{N} \cos \frac{l \pi}{N} \right| = \frac{1}{2} \left| \sin \frac{2 l \pi}{N} \right| = \frac{1}{4} |s_{2l}|.
\end{displaymath}
Combining all these computations we get
\begin{eqnarray}
  \pi_{res} \Big( \b G_4 & + & \frac{\a^2}{2} \{ G_3, F_3 \} \Big) \nonumber\\
 & = & \!\! \frac{1}{24N} \!\! \sum_{(k_1,k_2,k_3,k_4) \in K_4^{res}} \!\!\!\!\!\!\!\!\!\!\!\! (-1) (\b + \frac{3 \a^2}{2} c_{k_1 k_2 k_3 k_4}^S) \l_{k_1} \! \l_{k_2} \! \l_{k_3} \! \l_{k_4} \z_{k_1} \! \z_{k_2} \! \z_{k_3} \! \z_{k_4} \nonumber\\
& = & -\frac{4! \, (\b - \a^2)}{24N} \sum_{1 \leq l < \frac{N}{4}} \frac{s_{2l}}{4} \left( \z_l \z_{-N\!+\!l} \z_{\frac{N}{2}\!-\!l} \z_{-\frac{N}{2}\!-\!l} + \z_{-l} \z_{N\!-\!l} \z_{\frac{N}{2}\!+\!l} \z_{-\frac{N}{2}\!+\!l} \right) \nonumber\\
&& \underbrace{-\frac{3! \, (\b - \a^2)}{24N} \cdot \frac{2}{4} \left( \z_{\frac{N}{4}}^2 \z_{-\frac{3N}{4}}^2 + \z_{-\frac{N}{4}}^2 \z_{\frac{3N}{4}}^2 \right)}_{\textrm{only if }\frac{N}{4} \in \N} \label{reszetaforprov} \\
& = & -\frac{\b - \a^2}{4N} (R + R_\frac{N}{4}), \nonumber
\end{eqnarray}
with $R$ and $R_{\frac{N}{4}}$ as defined by (\ref{rresdef}) and (\ref{rn4def}), respectively. Hence Lemma \ref{resprojlemma} is proved.
\end{proof}
By Lemma \ref{resprojlemma}, if $N$ is even, equation (\ref{fkkkkeqn}) can be solved for any quadruple $(k,k',k'',k''') \in K_4 \setminus (K_4^N \cup K_4^{res})$ in such a way that $F^{(4)}_{\sigma(k,k',k'',k''')} = F^{(4)}_{(k,k',k'',k''')}$ for any permutation $\sigma(k,k',k'',k''')$ of $(k,k',k'',k''')$. With this choice of $F_4$ the canonical transformation $\Psi_2$ is then defined by $X_{F_4}^t|_{t=1}$. Composing $\Psi_1$ and $\Psi_2$, we obtain the transformation $\Xi := \Psi_1 \circ \Psi_2$ and have proved the following

\begin{prop} \label{localtheorem}
Assume that $N$ is even. The real analytic symplectic coordinate transformation $\z = \Xi(z)$, defined locally in a neighborhood of the origin $z=0$ in $\mathcal{Z}$, transforms the Hamiltonian $\hvt$ into the resonant Birkhoff normal form up to order $4$,
\begin{displaymath} 
  \hvt \circ \Xi = G_2 + \pi_N \left( \b G_4 + \frac{\a^2}{2} \{ G_3, F_3 \} \right) + \pi_{res} \left( \b G_4 + \frac{\a^2}{2} \{ G_3, F_3 \} \right) + O(z^5),
\end{displaymath}
with $G_2$, $\pi_N(\b G_4 + \frac{\a^2}{2} \{ G_3, F_3 \})$, and $\pi_{res} (\b G_4 + \frac{\a^2}{2} \{ G_3, F_3 \})$ given by (\ref{g2lambdadef}), (\ref{ping4ping3f3}), and (\ref{reszetaformula}), respectively.
\end{prop}

\begin{proof}[Proof of Theorem \ref{bnffputheoremeven}]
We start with the formula for $\hvt \circ \Xi$ given by Proposition \ref{localtheorem} and treat the normal form terms $G_2 + \pi_N(\b G_4 + \frac{\a^2}{2} \{ G_3, F_3 \})$ and the resonant non-normal form terms $\pi_{res} (\b G_4 + \frac{\a^2}{2} \{ G_3, F_3 \})$ separately. With the action variables $I = (I_k)_{\1N1}$ defined by (\ref{ikdef}) we see that $G_2 + \pi_N(\b G_4 + \frac{\a^2}{2} \{ G_3, F_3 \}) \! = \! H_{\a,\b}(I)$, where $H_{\a,\b}(I)$ is defined by (\ref{bnffpuodd}). Concerning $\pi_{res} (\! \b G_4 \! + \frac{\a^2}{2} \{ G_3, F_3 \})$, we first express it in terms of the real variables $(x_k, y_k)_{\1N1}$, related to the $\z_k$'s by $x_k = (\z_k + \z_{-k})/2$ and $y_k = (\z_{-k} - \z_k)/2i$. Note that
\begin{eqnarray}
 && \!\!\!\!\!\!\!\!\!\!\!\!\!\!\!\! \z_l \z_{-N+l} \z_{\frac{N}{2}-l} \z_{-\frac{N}{2}-l} + \z_{-l} \z_{N-l} \z_{-\frac{N}{2}+l} \z_{\frac{N}{2}+l} \nonumber\\
& = & 2 \; \textrm{Re} \, (\z_l \z_{-N+l} \z_{\frac{N}{2}-l} \z_{-\frac{N}{2}-l}) \nonumber\\
& = & \frac{1}{2} \Big( \left( x_l x_{N-l} + y_l y_{N-l} \right) \left( x_{\frac{N}{2} - l} x_{\frac{N}{2} + l} + y_{\frac{N}{2} - l} y_{\frac{N}{2} + l} \right) \nonumber\\
 && \quad - \left( x_l y_{N-l} - x_{N-l} y_l \right) \left( x_{\frac{N}{2} - l} y_{\frac{N}{2} + l} - x_{\frac{N}{2} + l} y_{\frac{N}{2} - l} \right) \Big) \nonumber\\
& = & 2 \left( J_l J_{\frac{N}{2} - l} - M_l M_{\frac{N}{2} - l} \right), \label{jjminusmm}
\end{eqnarray}
where for any $\1N1$
\begin{displaymath}
  J_k := \frac{1}{2} \left( x_k x_{N-k} + y_k y_{N-k} \right) \;\; \textrm{and} \;\; M_k := \frac{1}{2} \left( x_k y_{N-k} - x_{N-k} y_k \right).
\end{displaymath}
Hence $R$, given by (\ref{rresdef}), can be expressed in terms of $J_k$ and $M_k$ as follows
\begin{eqnarray}
R(J,M) & = & \sum_{1 \leq l < \frac{N}{4}} \!\! s_{2l} \left( \z_l \z_{-N+l} \z_{\frac{N}{2}-l} \z_{-\frac{N}{2}-l} + \z_{-l} \z_{N-l} \z_{\frac{N}{2}+l} \z_{-\frac{N}{2}+l} \right) \nonumber\\
& = & 4 \sum_{1 \leq l < \frac{N}{4}} \sin \frac{2 l \pi}{N} \left( J_l J_{\frac{N}{2} - l} - M_l M_{\frac{N}{2} - l} \right). \label{rformula}
\end{eqnarray}
Similarly, if $\frac{N}{4} \in \N$, one concludes from (\ref{jjminusmm}) that $R_\frac{N}{4}$, given by (\ref{rn4def}), can be expressed as
\begin{equation} \label{r4nformula}
  R_\frac{N}{4}(J,M) = \frac{1}{2} \left( \z_{\frac{N}{4}}^2 \z_{-\frac{3N}{4}}^2 + \z_{\frac{3N}{4}}^2 \z_{-\frac{N}{4}}^2 \right) = J_\frac{N}{4}^2 - M_\frac{N}{4}^2.
\end{equation}
Theorem \ref{bnffputheoremeven} now follows from the formulas (\ref{bnffpuodd}), (\ref{rformula}), and (\ref{r4nformula}).
\end{proof}

\section{Proof of Theorem \ref{bnfproperties}} \label{special}

To analyze the Hessian $Q_{\a, \b}$ of (\ref{bnffpuodd}) at $I=0$ we repeatedly encounter matrices of the form $E + \textrm{diag}(\mu_1, \ldots, \mu_{N-1})$, where $E$ is the $(N-1) \times (N-1)$-matrix
\begin{equation} \label{ematrixdef}
  E := \left( \begin{array}{ccc}
1 & \ldots & 1 \\
\vdots & & \vdots \\
1 & \ldots & 1
\end{array} \right)
\end{equation}
and $(\mu_k)_{\1N1}$ are given complex numbers. The determinant of the matrix $E + \textrm{diag}(\mu_1, \ldots, \mu_{N-1})$ can be explicitly computed.

\begin{lemma} \label{ematrreglemma}
Let $(\mu_k)_{1 \leq k \leq N-1}$ be given nonzero complex numbers. Then
\begin{equation} \label{detematrix}
  \det \left( E + \textrm{diag}(\mu_1, \ldots, \mu_{N-1}) \right) = \left( 1 + \sum_{k=1}^{N-1} \frac{1}{\mu_k} \right) \cdot \prod_{k=1}^{N-1} \mu_k.
\end{equation}
In particular, $E + \textrm{diag}(\mu_1, \ldots, \mu_{N-1})$ is regular if and only if $\sum_{k=1}^{N-1} \frac{1}{\mu_k} \neq -1$.
\end{lemma}

\begin{proof}
Expanding $\det (E + \textrm{diag}(\mu_1, \ldots, \mu_{N-1}))$ with respect to its rows it follows that
\begin{displaymath}
  \det (E + \textrm{diag}(\mu_1, \ldots, \mu_{N-1})) = \prod_{k=1}^{N-1} \mu_k + \sum_{k=1}^{N-1} \prod_{l \neq k} \mu_l.
\end{displaymath}
This leads to formula (\ref{detematrix}).
\end{proof}

First let us treat the case $\a = 0$, $\b \neq 0$. Using the notation introduced in section \ref{lbc}, one obtains the following proposition. It improves earlier results of Rink \cite{rink01}.

\begin{prop} \label{bchaincor}
Let $N$ be odd and assume that $\a = 0$ in (\ref{potentialdef}). Then the following holds:
  \begin{itemize}
  \item[(i)] The Birkhoff normal form of $H_V$ up to order $4$ is given by $\frac{N P^2}{2} + \H0b(I)$ where
    \begin{equation} \label{bnfbchain}
        \H0b(I) = 2 \sum_{k=1}^{N-1} \l_k^2 I_k + \frac{\b}{4N} \left( \sum_{k=1}^{N-1} \l_k^4 I_k^2 + 2 \!\!\!\! \sum_{1 \leq l \neq m \leq N-1} \!\!\!\! \l_l^2 \l_m^2 I_l I_m \right).
    \end{equation}
  \item[(ii)] For any $\b \neq 0$, $\H0b(I)$ is nondegenerate at $I=0$.
  \end{itemize}
\end{prop}

\begin{proof}
The Birkhoff normal form (\ref{bnfbchain}) of $H_V$ is given by the formula (\ref{bnffpuodd}) evaluated at $\a = 0$. To investigate the Hessian $Q_{0,\b}$ of $\H0b(I)$ at $I=0$ we write
\begin{equation} \label{q0bdecomp}
  Q_{0,\b} = \frac{\b}{4N} \, \Delta \, P \, \Delta,
\end{equation}
where
\begin{equation} \label{deltamatrixdef}
\Delta = \textrm{diag} \, \left( \sin\frac{k\pi}{N} \right)_{\1N1}
\end{equation}
and
\begin{displaymath}
P = 2 \cdot \left( E - \frac{1}{2} \, \textrm{Id}_{N-1} \right).
\end{displaymath}
In view of (\ref{detematrix}) it follows that
\begin{displaymath}
  \det Q_{0,\b} = \left( \frac{\b}{4N} \right)^{N-1} \cdot \det P \cdot \prod_{k=1}^{N-1} \sin^2 \frac{k\pi}{N}
\end{displaymath}
where by Lemma \ref{ematrreglemma},
\begin{displaymath}
  \det P = 2^{N-1} \left(1 - 2(N-1) \right) (-1/2)^{N-1} = (-1)^N (2N-3) \neq 0.
\end{displaymath}
Hence, if $\b \neq 0$, $\det Q_{0,\b} \neq 0$,
and the nondegeneracy of $\H0b(I)$ at $I=0$ follows.
\end{proof}

\begin{lemma} \label{signbchain}
If $\b < 0$, then $Q_{0,\b}$ has one negative eigenvalue, whereas if $\b > 0$, then $Q_{0,\b}$ has $N-2$ negative eigenvalues. In particular, for any $\b \in \R \setminus \{ 0 \}$, $Q_{0,\b}$ is indefinite (and $\H0b$ is therefore not convex).
\end{lemma}

\begin{proof}
We want to use the decomposition (\ref{q0bdecomp}) of $Q_{0,\b}$ to show that $Q_{0,\b}$ can be deformed continuously to $\frac{\b}{4N} P$: Consider for $0 \leq t \leq 1$
\begin{displaymath}
  Q_{0,\b}(t) := \frac{\b}{4N} (t \, \Delta + (1-t) \, \textrm{Id}) \; P \; (t \, \Delta + (1-t) \, \textrm{Id}).
\end{displaymath}
As $t \, \Delta + (1-t) \, \textrm{Id}$ is positive definite for any $0 \leq t \leq 1$ and $P$ is regular, $Q_{0,\b}(t)$ is a symmetric regular $(N-1) \times (N-1)$-matrix. For $t=0$, $Q_{0,\b}(0) = \frac{\b}{4N} P$, whereas for $t=1$, $Q_{0,\b}(1) = Q_{0,\b}$. Therefore, index$(Q_{0,\b})$ (i.e. the number of negative eigenvalues of $Q_{0,\b}$) coincides with index$(\frac{\b}{4N} P)$. The eigenvalues of $P$ are $\mu_1 = 2N-3$ with multiplicity one and $\mu_2 = -1$ with multiplicity $N-2$.
\end{proof}

We now turn to the case $\a \neq 0$.
\begin{prop} \label{chaingenproperties}
Assume that $\a \neq 0$ in (\ref{potentialdef}). Then, for $\a$ fixed, $\det Q_{\a,\b}$ is a polynomial in $\b$ of degree $N-1$ and has $N-1$ real zeroes (counted with multiplicities) which we list in increasing order and denote by $\b_k = \b_k(\a)$ ($\1N1$). They satisfy $0 < \b_1 < \a^2 < \b_2 \leq \ldots \leq \b_{N-1}$ and contain the $\llcorner (N-1)/2 \lrcorner$ distinct numbers
\begin{displaymath}
  \a^2 \left( 1 + \left( \sin^2 \frac{k \pi}{N} \right)^{-1} \right) \quad (1 \leq k \leq \llcorner \frac{N-1}{2} \lrcorner).
\end{displaymath}
Moreover
\begin{displaymath}
  \textrm{index} \, (\Qab) = \left\{  \begin{array}{ll}
1 & \quad \textrm{for } \b < \b_1 \\
0 & \quad \textrm{for } \b_1 < \b < \b_2 \\
N-2 & \quad \textrm{for } \b > \b_{N-1}
\end{array} \right.
\end{displaymath}
\end{prop}

\begin{proof}
Fix $\a \in \R \setminus \{ 0 \}$ and consider the map $\b \mapsto \det (Q_{\a, \b})$. It follows from (\ref{bnffpuodd}) that $\det (Q_{\a, \b})$ is a polynomial in $\b$ of degree at most $N-1$,
\begin{displaymath}
  \det (Q_{\a, \b}) = \sum_{j=0}^{N-1} q_j \b^j,
\end{displaymath}
where $q_0 = \det(Q_{\a,0})$ and $q_{N-1} = \det(Q_{0,1})$. By Proposition \ref{bchaincor}, $\det (Q_{0,1}) \neq 0$, hence the degree of the polynomial $\det (Q_{\a, \b})$ is $N-1$. We claim that $\det(\Qab)$ has at least $N-1$ real zeroes (counted with multiplicities). For $|\b|$ large enough, index$(\Qab)$ is equal to index$(Q_{0,\b})$. By Lemma \ref{signbchain}, index$(Q_{0,\b})$ is $N-2$ for $\b > 0$ and $1$ for $\b < 0$. Hence there exists $R > 0$ such that index$(\Qab) = N-2$ for any $\b > R$ and index$(\Qab) = 1$ for any $\b < -R$. For $\b = \a^2$, $Q_{\a,\a^2}$ is a positive multiple of the identity matrix, hence index$(Q_{\a,\a^2}) = 0$. It then follows that index$(\Qab)$ must change at least once in the open interval $(-\infty, \a^2)$ and at least $N-2$ times (counted with multiplicities) in $(\a^2, \infty)$. Since a change of index$(Q_{\a, \b})$ induces a zero of $\det (\Qab)$ (counted with multiplicities), our consideration shows that $\b \mapsto \det (\Qab)$ has at least $N-1$ real zeroes. Thus $\b \mapsto \det (\Qab)$ has precisely $N-1$ real zeroes and we have $\b_1(\a) < \a^2 < \b_2(\a)$.

Next we prove that $\b_1(\a) > 0$, i.e. that $\Qab$ is regular for any $\b \leq 0$. Write $\Qab$ as a product,
\begin{displaymath}
  Q_{\a,\b} = \frac{\a^2-\b}{4N} \, \Delta \, \Pab \, \Delta,
\end{displaymath}
where $\Delta$ is given by (\ref{deltamatrixdef}) and $\Pab$ is given by
\begin{displaymath}
\Pab = -2 \left( E + \textrm{diag} \, \left( -\frac{1}{2} \left( 1 + \frac{\gab}{\sin^2\frac{k\pi}{N}} \right) \right)_{\1N1} \right), \label{pabspecial}
\end{displaymath}
where $E$ is given by (\ref{ematrixdef}) and
\begin{displaymath}
  \gab := \frac{\a^2}{\a^2-\b}.
\end{displaymath}
As $-\infty < \b \leq 0$ it follows that $0 < \gab \leq 1$ and $\mu_k = -\frac{1}{2} \left( 1 + \frac{\gab}{\sin^2\frac{k\pi}{N}} \right) < 0$ for any $\1N1$. Lemma \ref{ematrreglemma} says that $\Pab$ is regular if $f(\gab) \neq 0$ where
\begin{displaymath}
  f(\g) := 1 - 2 \sum_{k=1}^{N-1} \left( 1 + \g / \sin^2 \frac{k \pi}{N} \right)^{-1}
\end{displaymath}
in the interval $0 < \g \leq 1$. Note that $f(\g)$ is increasing in $0 < \g \leq 1$ and $f(1)$ can be estimated as follows. 
Using that $N$ is assumed to be odd one has
\begin{eqnarray*}
  f(1) & = & 1 - 4 \sum_{k=1}^{\frac{N-1}{2}} \frac{\sin^2\frac{k\pi}{N}}{1 + \sin^2\frac{k\pi}{N}} < 1 - 4 \frac{\sin^2 \frac{(N-1)\pi}{2N}}{1 + \sin^2 \frac{(N-1)\pi}{2N}} \\
&& = 1 - 4 \frac{\cos^2 \frac{\pi}{2N}}{1 + \cos^2 \frac{\pi}{2N}} = -3 + \frac{4}{1 + \cos^2 \frac{\pi}{2N}}.
\end{eqnarray*}
As for $N \geq 3$
\begin{displaymath}
  -3 + \frac{4}{1 + \cos^2 \frac{\pi}{2N}} < -3 + \frac{4}{1 + \cos^2 \frac{\pi}{6}} = -\frac{5}{7}
\end{displaymath}
we conclude that $f(1) < 0$. Hence we have shown that $f(\g) < 0$ for $0 < \g \leq 1$, and therefore $\Pab$ is regular for $\b \leq 0$ by Lemma \ref{ematrreglemma}. Hence we have proved that $0 < \b_1(\a)$.

Finally introduce $\mu_k := -\frac{1}{2} (1 + \gab/\sin^2 \frac{k \pi}{N})$ and note that for $\b$ with $\gab = -\sin^2 \frac{k_0 \pi}{N}$ for some $1 \leq k_0 \leq \llcorner \frac{N-1}{2} \lrcorner$ one has $\mu_{k_0} = \mu_{N-k_0} = 0$. As $k_0 \neq N - k_0$ if $1 \leq k_0 \leq \llcorner (N-1)/2 \lrcorner$ it then follows that $\Pab$ has two equal rows and is therefore singular. Note that $\gab = -\sin^2 \frac{k_0 \pi}{N}$ corresponds to $\b = \a^2 (1 + \sin^{-2} \frac{k_0 \pi}{N})$ and we have proved that $\b \mapsto \det (\Qab)$ has at least $\llcorner (N-1)/2 \lrcorner$ different zeroes in the interval $(\a^2, \infty)$. The statement about index$(\Qab)$ easily follows from the above analysis.
\end{proof}

\begin{proof}[Proof of Theorem \ref{bnfproperties}]
Part (i) is proved by Proposition \ref{chaingenproperties}, whereas (ii) follows from Proposition \ref{bchaincor} and Lemma \ref{signbchain}.
\end{proof}

\appendix

\section{Proof of Lemma \ref{nonres4lemma}} \label{nonresapp}

For the convenience of the reader, we provide a detailed proof of Lemma \ref{nonres4lemma} in this appendix. This lemma and its proof are due to Beukers and Rink - see (\cite{rink01}, Appendix A). Recall that $K_4 \setminus K_4^N \subseteq \Z^4$ denotes the subset of quadruples $(k_1,k_2,k_3,k_4)$ satisfying $1 \leq |k_i| \leq N-1$ ($1 \leq i \leq 4$) and $k_1 + k_2 + k_3 + k_4 \equiv 0$ mod $N$ so that there are no integers $l,m$ with $\{ l,m,-l,-m \} = \{ k_1,k_2,k_3,k_4 \}$, and
\begin{displaymath}
  K_4^{res} := K_{res}^+ \cup K_{res}^- \subseteq K_4
\end{displaymath}
where
\setlength\arraycolsep{2pt}{
\begin{eqnarray*} 
   K_{res}^\pm & := & \Big\{ (k_1,k_2,k_3,k_4) \in K_4 | \; \exists \, l \in \N: 1 \leq l \leq \frac{N}{4} \;\; \textrm{so that} \\
&& \quad \{ k_1,k_2,k_3,k_4 \} = \{ \pm l, \pm l \mp N, \frac{N}{2} \mp l,-\frac{N}{2} \mp l \} \Big\}.
\end{eqnarray*}}
Note that $K_4^{res} = \emptyset$ if $N$ is odd. Let us restate Lemma \ref{nonres4lemma} as follows:

\begin{lemma} \label{nonreslemmaapp}
(\cite{rink01}) Let $(k_1,k_2,k_3,k_4)$ be an element of $K_4 \setminus K_4^N$. Then $(k_1,k_2,k_3,k_4) \in K_4^{res}$ if and only if
\begin{displaymath}
  \sin \frac{k_1 \pi}{N} + \sin \frac{k_2 \pi}{N} + \sin \frac{k_3 \pi}{N} + \sin \frac{k_4 \pi}{N} = 0.
\end{displaymath}
\end{lemma}

Let us make a few preparations for the proof of Lemma \ref{nonreslemmaapp}. By a straightforward computation one sees that the ``only if''-part of the claimed equivalence holds:

\begin{lemma} \label{nonreslemmaifpart}
For any $(k_1,k_2,k_3,k_4) \in K_4^{res}$, one has $\sum_{i=1}^4 \sin \frac{k_i \pi}{N} = 0$.
\end{lemma}

So it remains to prove the converse. First we consider some special cases.

\begin{lemma} \label{simplecases}
Let $(k_1,k_2,k_3,k_4) \in K_4 \setminus (K_4^N \cup K_4^{res})$. If there exist $l,m,n \in \Z$ such that
\begin{itemize}
\item[(i)]  $\{ k_1,k_2,k_3,k_4 \} = \{ l,-l,m,n \}$, or
\item[(ii)] $\{ k_1,k_2,k_3,k_4 \} = \{ l,N-l,m,n \}$ with $1 \leq l \leq N-1$, or
\item[(iii)] $\{ k_1,k_2,k_3,k_4 \} = \{ l,-N-l,m,n \}$ with $-(N-1) \leq l \leq -1$,
\end{itemize}
then
\begin{displaymath}
  \sum_{i=1}^4 \sin \frac{k_i \pi}{N}\neq 0.
\end{displaymath}
\end{lemma}

\begin{proof}
In case (i), it follows that $m+n = N$ (and thus $1 \leq m,n \leq N-1$) or $m+n = -N$ (and thus $-(N-1) \leq m,n \leq -1$). Hence in both cases, $\sm$ and $\sn$ have the same sign and $\sum_{i=1}^4 \sin \frac{k_i \pi}{N} = \sm + \sn \neq 0$. In the case (ii), by assumption, $m+n \equiv 0$ mod $N$. The case $m+n = 0$ has already been treated under (i). If $m+n = N$, then $\sin \frac{k_i \pi}{N} > 0$ for any $1 \leq i \leq 4$. If $m+n = -N$, then $m < 0$, and $m \notin \{ -l,-N+l \}$. Thus $n = -N-m < 0$ and therefore $\sum_{i=1}^4 \sin \frac{k_i \pi}{N} = 2 \sl - 2 \sin \frac{(-m) \pi}{N} \neq 0$. The case (iii) is treated similarly as (ii).
\end{proof}

Another special case in treated in the following lemma.
\begin{lemma} \label{nonressimple2}
Assume that $(k_1,k_2,k_3,k_4) \in K_4 \setminus K_4^N$ satisfies
\begin{equation} \label{k1kineq0}
  k_i + k_j \not \equiv 0 \textrm{ mod } N \quad \forall \, 1 \leq i, j \leq 4.
\end{equation}
If there exist $l,n \in \{ k_1,k_2,k_3,k_4 \}$ with
\begin{equation} \label{sinkikj}
  \sin \frac{l \pi}{N} + \sin \frac{n \pi}{N} = 0,
\end{equation}
then
\begin{equation} \label{reseqn}
 \sum_{i=1}^4 \sin \frac{k_i \pi}{N} = 0
\end{equation}
implies that $(k_1,k_2,k_3,k_4) \in K_4^{res}$.
\end{lemma}

\begin{proof}
From the assumptions (\ref{k1kineq0})-(\ref{sinkikj}) it follows that there exists $1 \leq l \leq N-1$ so that $\{ k_1,k_2,k_3,k_4 \} = \{ l,-N+l,m,n \}$ for some $m, n \in \Z$. Then $\sl + \sin \frac{(-N+l) \pi}{N} = 0$ and hence by (\ref{reseqn}), $\sm + \sn = 0$. W.l.o.g. assume that $1 \leq m \leq N-1$. Then either $n=-m$ or $n=-N+m$. If $n=-m$, then $(k_1,k_2,k_3,k_4) \in K_4^{res}$ by Lemma \ref{simplecases} (i). If $n=-N+m$, then one has
\begin{displaymath}
  \sum_{i=1}^4 k_i = 2l-N+2m-N = 2(l+m) - 2N.
\end{displaymath}
Note that $2(l+m) - 2N$ cannot be an even multiple of $N$, as otherwise $l+m \equiv 0$ mod $N$, violating (\ref{k1kineq0}). If, in addition, $N$ is odd, then $2(l+m) - 2N$ cannot be odd multiple of $N$. Hence in the case $N$ is odd we conclude that $\sum_{i=1}^4 k_i \not \equiv 0$ mod $N$, contradicting the assumption $(k_1,k_2,k_3,k_4) \in K_4$.

If $N$ is even, it is however possible that $2(l+m) - 2N$ equals $\pm N$: If $2(l+m) - 2N = N$, i.e. $l+m = \frac{3}{2}N$, it follows that $\frac{N}{2} < l,m \leq N-1$, and $(k_1,k_2,k_3,k_4) \in K_{res}^-$ with $l'=l-\frac{N}{2}$. If $2(l+m) - 2N = -N$, i.e. $l+m = \frac{N}{2}$, it follows similarly that $(k_1,k_2,k_3,k_4) \in K_{res}^+$ with $l'=l$. In both cases, we conclude that $(k_1,k_2,k_3,k_4) \in K_4^{res}$.
\end{proof}

In view of Lemma \ref{simplecases} and Lemma \ref{nonressimple2} in order to prove Lemma \ref{nonreslemmaapp} it remains to show the following

\begin{lemma} \label{nonresessential}
Assume that $(k_1,k_2,k_3,k_4) \in K_4$ satisfies (\ref{k1kineq0}). If for any $1 \ \leq i,j \leq 4$
\begin{equation} \label{slsmneq0}
  \sin \frac{k_i \pi}{N} + \sin \frac{k_j \pi}{N} \neq 0.
\end{equation}
(and thus $(k_1,k_2,k_3,k_4) \notin K_4^N \cup K_4^{res}$), then
\begin{displaymath}
  \sum_{i=1}^4 \sin \frac{k_i \pi}{N} \neq 0.
\end{displaymath}
\end{lemma}

To prove Lemma \ref{nonresessential} let us first rewrite (\ref{reseqn}), using Euler's formula for the sine function,
\begin{equation} \label{zetaeqn0}
\sum_{1 \leq |j| \leq 4} \zeta_j = 0
\end{equation}
where $\zeta_{\pm j} = \pm e^{\pm i k_j \pi / N}$ are $2N$'th roots of unity. Note that for any quadruple $(k_1,k_2,k_3,k_4) \in K_4 \setminus K_4^N$ satisfying (\ref{slsmneq0}) one has
\begin{displaymath}
\z_j + \z_{j'} \neq 0 \quad \forall \, 1 \leq |j| \leq |j'| \leq 4
\end{displaymath}
Indeed for any $1 \leq |j| \leq |j'| \leq 4$ one has Im $\z_j +$ Im $\z_{j'} = \sin \frac{k_{|j|} \pi}{N} + \sin \frac{k_{|j'|} \pi}{N}$ which does not vanish by assumption (\ref{slsmneq0}).

Let us first discuss equation (\ref{zetaeqn0}) and its solutions in general. For convenience let us rewrite (\ref{zetaeqn0}) as
\begin{equation} \label{zetaeqn}
  \z_1 + \ldots + \z_8 = 0.
\end{equation}
We are interested in the solutions $(\z_l)_{1 \leq l \leq 8}$ of the equation (\ref{zetaeqn}) on the unit circle $S^1 := \{ z \in \C \big| |z| = 1 \}$.

We need an auxiliary result which we want to discuss first. Let $n \geq 2$ be arbitrary and assume that the sequence $(\z_i)_{1 \leq i \leq n} \subseteq S^1$ has no vanishing subsums (i.e. $\sum_{l \in J} \z_l \neq 0$ for any $\emptyset \neq J \subsetneq \{ 1, \ldots, n \}$) and satisfies the equation
\begin{equation} \label{zetaeqngen}
\sum_{i=1}^n \z_i = 0.
\end{equation}
Let $M \in \N$ be the smallest integer with the property that $(\z_i / \z_j)^M = 1$ for all $1 \leq i,j \leq n$. Then there exists $\xi \in S^1$ so that $\z_i^M = \xi^M$ for any $1 \leq i \leq n$. W.l.o.g. we can assume that $\xi = 1$. Furthermore, let $p^k$ be a prime power dividing $M$ so that $M/p^k$ and $p$ are relatively prime and define
\begin{equation} \label{mprimedef}
  M' =: M/p \;\; \textrm{and} \;\; \eta := e^{2 \pi i / p^k}. 
\end{equation}
Then for any $1 \leq l \leq n$ there exists a unique integer $0 \leq \mu(l) \leq p-1$ such that $\z_l = \tilde{\z}_l \cdot \eta^{\mu(l)}$ where $\tilde{\z}_l$ is an element of the field $K := \Q(e^{2 \pi i / M'})$. (As $\z_l^M = 1$ there exists $0 \leq r_l \leq M-1$ with $\z_l = e^{\frac{2 \pi i}{M} r_l}$. If $r_l \equiv 0$ mod $p$ choose $\mu(l) = 0$. If $r_l \not \equiv 0$ mod $p$ choose $1 \leq \mu(l) \leq p-1$ so that $r_l \equiv \frac{M}{p^k} \mu(l)$ mod $p$.) Hence (\ref{zetaeqngen}) can be written as
\begin{equation} \label{polannzeta}
  0 = \sum_{l=1}^n \zeta_l = \sum_{l=1}^n \tilde{\z}_l \eta^{\mu(l)} = \sum_{s=0}^{p-1} \left( \sum_{l \in \mu^{-1}(s)} \tilde{\z}_l \right) \eta^s
\end{equation}
We need the following algebraic fact (see e.g. \cite{vdw}, \S 60-61):
\begin{prop} \label{algprop}
The minimal polynomial of $\eta = e^{2 \pi i / p^k}$ over the field $K = \Q(e^{2 \pi i / M'})$ is given by $X^p - \eta^p$ if $k \geq 2$ and $X^{p-1} + X^{p-2} + \ldots + X + 1$ if $k=1$.
\end{prop}
We now claim that $M$ is square-free, or equivalently that for any prime power $p^k$ dividing $M$,
\begin{equation} \label{sqfr}
  k = 1.
\end{equation}
Indeed, equation (\ref{polannzeta}) shows that the minimal polynomial of $\z$ has degree at most $p-1$, which by Proposition \ref{algprop} is only satisfied in the case $k=1$.

Further we claim that there exists $\sigma \in \C \setminus \{ 0 \}$ so that
\begin{equation} \label{sigmaexists}
  \sum_{l \in \mu^{-1}(s)} \tilde{\z}_l = \sigma \quad \forall \, 0 \leq s \leq p-1.
\end{equation}
The existence of such a $\sigma$ follows from Proposition \ref{algprop}: As $k = 1$ by (\ref{sqfr}), the minimal polynomial of $\eta$ over $K$ is given by $X^{p-1} + X^{p-2} + \ldots + X + 1$. Since this is a polynomial of degree $p-1$ the polynomial on the right hand side of (\ref{polannzeta}) must be a scalar multiple of the minimal polynomial. Hence all the coefficients $\sum_{l \in \mu^{-1}(s)} \tilde{\z}_l$ have the same value $\sigma \in \C$. As $\sum_{l \in \mu^{-1}(s)} \z_l = \sigma \eta^s$ the additional property $\sigma \neq 0$ follows from the assumption that there are no vanishing subsums. Hence we can assume w.l.o.g. that $\sigma = 1$.

Next we claim that
\begin{equation} \label{pleqn}
  p \leq n.
\end{equation}
In other words, possible prime factors of $M$ are bounded by the number of summands in (\ref{zetaeqngen}). To prove (\ref{pleqn}), note that it follows from (\ref{sigmaexists}) that for any $0 \leq s \leq p-1$ there exists $1 \leq l \leq n$ such that $\mu(l) = s$, i.e. the map $\mu: \{ 1, \ldots, n\} \to \{ 0, \ldots, p-1 \}$ is onto. This establishes (\ref{pleqn}).

The map $\mu$ induces the \emph{partition} $(\sharp \, \mu^{-1}(s))_{0 \leq s \leq p-1}$ of the positive integer $n$ into $p$ summands,
\begin{equation}
  n=\sum_{s=0}^{p-1} \sharp \, \mu^{-1}(s)
\end{equation}

\begin{lemma} \label{applemma1}
(\cite{rink01}, Appendix A)
For any solution $\{ \z_1, \ldots, \z_8 \}$ of (\ref{zetaeqn}) contained in $S^1$ without vanishing subsums there exists $\xi \in S^1$ such that either
\begin{equation} \label{rofsol1}
  \{ \z_1, \ldots, \z_8 \} = \{ -\xi \a, -\xi \a^2 \} \cup \{ \xi \g^j \, | \, 1 \leq j \leq 6 \}
\end{equation}
or
\begin{equation} \label{rofsol2}
  \{ \z_1, \ldots, \z_8 \} = \{ -\xi \a^l, -\xi \a^l \cdot \b^i, -\xi \a^l \cdot \b^j \, | \, 1 \leq l \leq 2 \} \cup \{ \xi \b^k, \xi \b^m \},
\end{equation}
where the quadruple $(i,j,k,l)$ is a permutation of $(1,2,3,4)$ and
\begin{displaymath}
  \a := e^\frac{2 \pi i}{3}, \quad \b := e^\frac{2 \pi i}{5}, \quad \g := e^\frac{2 \pi i}{7}.
\end{displaymath}
\end{lemma}

\begin{proof}
By a straightforward computation one verifies that the sets of the form (\ref{rofsol1}) or (\ref{rofsol2}) satisfy (\ref{zetaeqn}). It remains to prove that these are the only solutions of (\ref{zetaeqn}) of this type.

We classify the solutions of (\ref{zetaeqn}) according to the possible values of $p$, which we now assume to be the largest prime dividing $M$. Since $n=8$, by (\ref{pleqn}), the possible values of $p$ are $2$, $3$, $5$, and $7$. If $p=2$, then, by (\ref{sqfr}), $M=2$ and therefore there exists $\xi \in S^1$ so that $\z_j = \pm \xi$ for any $1 \leq j \leq n$. In this case there exists a solution of (\ref{zetaeqngen}) without vanishing subsums only if $n=2$. (In this case, they are given by $\{ \z_1, \z_2 \} = \xi \{ 1, -1 \}$ with $\xi \in S^1$.) If $p=3$, then $M=3$ or $M=3 \cdot 2$, and there exists a solution of (\ref{zetaeqngen}) without vanishing subsums only if $n=3$. (In this case, they are given by $\{ \z_1, \z_2, \z_3 \} = \xi \{ 1, \a, \a^2 \}$ with $\xi \in S^1$.) If $p=5$, then $\eta = \b$ in (\ref{mprimedef}). Up to permutations, there are the following three partitions of $8$ into $5$ summands, $(4,1,1,1,1)$, $(3,2,1,1,1)$, and $(2,2,2,1,1)$. In a straightforward way one shows that the partitions $(4,1,1,1,1)$ and $(3,2,1,1,1)$ and their permutations give rise to solutions of the equation (\ref{zetaeqn}) with vanishing subsums. E.g. the solutions corresponding to $(4,1,1,1,1)$ are given by $\xi \cdot (-\b, -\b^2, -\b^3, -\b^4, \b, \b^2, \b^3, \b^4)$ with $\xi \in S^1$, whereas the solutions corresponding to $(3,2,1,1,1)$ are $\xi \cdot (-i, 1, i, -\a \b, -\a^2 \b, \b^2, \b^3, \b^4)$ with $\xi \in S^1$. On the other hand the partition $(2,2,2,1,1)$ leads to the solutions
\begin{displaymath}
  (\z_1, \ldots, \z_8) = \xi (-\a, -\a^2, -\a \b, -\a^2 \b, -\a \b^2, -\a^2 \b^2, \b^3, \b^4)
\end{displaymath}
with $\xi \in S^1$. They are the solutions (\ref{rofsol2}) with $(i,j,k,m) = (1,2,3,4)$. Permutations of the partition $(2,2,2,1,1)$ again lead to solutions of the type (\ref{rofsol2}), but with $(i,j,k,m)$ given by a permutation of $(1,2,3,4)$.

If $p=7$, then $\eta = \g$ in (\ref{mprimedef}). Then, up to permutations, $(2,1,1,1,1,1,1)$ is the only possible partition of $8$ into $7$ summands. The partition $(2,1,1,1,1,1,1)$ leads to the solutions
\begin{displaymath}
  (\z_1, \ldots, \z_8) = \xi (-\a, -\a^2, \g, \ldots, \g^6)
\end{displaymath}
with $\xi \in S^1$, where we used that $1 = -\a - \a^2$. They are of type (\ref{rofsol1}). Any permutation of $(2,1,1,1,1,1,1)$ leads to the same kind of solutions.
\end{proof}

\begin{lemma} \label{applemma2}
(\cite{rink01}, Appendix A) For any solution $\{ \z_1, \ldots, \z_8 \}$ of (\ref{zetaeqn}) contained in $S^1$ without vanishing subsums of length $2$ but having a vanishing subsum of length $3$, $4$, or $5$, there exist $\xi$, $\xi' \in S^1$ such that
\begin{equation} \label{rofsol3}
  \{ \z_1, \ldots, \z_8 \} = \{ \xi \a^l | 0 \leq l \leq 2 \} \cup \{ \xi' \b^m| 0 \leq m \leq 4 \},
\end{equation}
where again $\a = e^{2 \pi i / 3}$ and $\b = e^{2 \pi i / 5}$.
\end{lemma}

\begin{proof}
Again, one verifies by a direct computation that the solutions (\ref{rofsol3}) of (\ref{zetaeqn}) have the desired properties. It remains to prove that they are the only ones. First note that under the hypotheses of the lemma, vanishing subsums of length $4$ cannot occur, since the latter ones would imply the existence of vanishing subsums of length $2$, which by assumption is excluded. Hence, in order to find solutions of (\ref{zetaeqngen}) for $n=8$ with the desired properties, we have to find all solutions of (\ref{zetaeqngen}) without vanishing subsums for $n=3$ and $n=5$. Note that by (\ref{pleqn}), $p=n$ for $n=3$ or $n=5$. By the considerations in the proof of Lemma \ref{applemma1}, the former ones are given by $(\z_1, \z_2, \z_3) = \xi (1, \a, \a^2)$ and the latter ones by $(\z_1, \ldots, \z_5) = \xi' (1, \b, \b^2, \b^3, \b^4)$ with $\xi, \xi' \in S^1$. This proves the lemma.
\end{proof}

We are now ready to prove Lemma \ref{nonresessential}.
\begin{proof}[Proof of Lemma \ref{nonresessential}]
We first select from (\ref{rofsol1}), (\ref{rofsol2}) and (\ref{rofsol3}) all the solutions $(\z_1, \ldots, \z_8)$ of (\ref{zetaeqn}) which are of the form (\ref{reseqn}) (after multiplication by $2i$). This amounts to selecting the solutions $(\z_1, \ldots, \z_8)$ of (\ref{zetaeqn}) having the property that $\{ \z_1, \ldots, \z_8 \}$ is invariant under the map $\z \mapsto -\z^{-1}$. It requires to choose $\xi$ and $\xi'$ in (\ref{rofsol1}), (\ref{rofsol2}), and (\ref{rofsol3}) appropriately. Let us explain this procedure in detail for the solutions of type (\ref{rofsol1}).

First we rewrite the solution (\ref{rofsol1}),
\begin{displaymath}
(\z_1, \ldots, \z_8) = \xi \cdot (-\a, -\a^2, \g, \g^2, \g^3, \g^4, \g^5, \g^6) = e^\frac{2 \pi i x}{42} \left( e^\frac{2 \pi i t_k}{42} \right)_{1 \leq k \leq 8},
\end{displaymath}
where
\begin{equation} \label{t1t8mod42}
  (t_1, \ldots, t_8) = (6,7,12,18,24,30,35,36) \quad \textrm{and} \quad x \in \R / 42\Z.
\end{equation}
The required invariance of the set of the $\z_k$'s under the map $\z \mapsto -\z^{-1}$ is equivalent to the invariance of the set of the $(t_k+x)$'s under the map $t \mapsto 21-t$ (mod $42$). 
Since the set (\ref{t1t8mod42}) of the $t_k$'s is invariant under the map $t \mapsto -t$ (mod $42$), $\{ t_k + x | 1 \leq k \leq 8 \}$ is invariant under $t \mapsto 21 - t$ (mod $42$), if we choose $x := \frac{21}{2}$ or $\xi = i$. Then the equation $\sum_{i=1}^8 \z_i = 0$ reads
\begin{displaymath}
  e^\frac{11 \pi i}{14} + e^\frac{5 \pi i}{6} + e^\frac{15 \pi i}{14} + e^\frac{19 \pi i}{14} + e^\frac{23 \pi i}{14} + e^\frac{27 \pi i}{14} + e^\frac{\pi i}{6} + e^\frac{3 \pi i}{14} = 0,
\end{displaymath}
or $\sin \frac{\pi}{6} + \sin \frac{3\pi}{14} + \sin \frac{15\pi}{14} + \sin \frac{19\pi}{14} = 0$. Choosing all arguments in $(0, \pi)$, the latter identity reads
\begin{equation} \label{sinres1}
  \sin \frac{\pi}{6} + \sin \frac{3\pi}{14} - \sin \frac{\pi}{14} - \sin \frac{5\pi}{14} = 0.
\end{equation}

For the solutions of type (\ref{rofsol2}), one gets
\begin{equation} \label{sinres2}
  \sin \frac{\pi}{6} + \sin \frac{13\pi}{30} - \sin \frac{7\pi}{30} - \sin \frac{3\pi}{10} = 0
\end{equation}
and
\begin{equation} \label{sinres3}
  \sin \frac{\pi}{6} + \sin \frac{\pi}{30} - \sin \frac{11\pi}{30} + \sin \frac{\pi}{10} = 0.
\end{equation}
Let us briefly explain how (\ref{sinres2})-(\ref{sinres3}) can be obtained. Note that from the $24$ permutations of $(1,2,3,4)$ in (\ref{rofsol2}), there are only six which lead to different sets of the $\z_i$'s, since interchanging $i$ and $j$ or $k$ and $m$ leaves the set on the right hand side of (\ref{rofsol2}) invariant. In the resulting six different cases, we again write $\{ \z_1, \ldots, \z_8 \} = \xi \cdot \{ e^{2 \pi i \cdot \frac{t_1}{30}}, \ldots, e^{2 \pi i \cdot \frac{t_8}{30}} \}$ with $t_i$ in $\R / 30\Z$. Then, up to translations, there are only two different types of solutions emerging from these six cases. With the appropriate choices of $\xi$, one gets the solutions (\ref{sinres2}) and
(\ref{sinres3}).

Finally, for the solutions of type (\ref{rofsol3}), one gets
\begin{equation} \label{sinres4}
  \sin \frac{\pi}{2} - \sin \frac{\pi}{6} + \sin \frac{\pi}{10} - \sin \frac{3\pi}{10} = 0.
\end{equation}
The procedure to obtain (\ref{sinres4}) is basically the same as in the preceding cases. We write (\ref{rofsol3}) as $\{ \z_1, \ldots, \z_8 \} = \xi \cdot \{ \a^l, \; \l \cdot \b^m | 0 \leq l \leq 2, \; 0 \leq m \leq 4 \}$ and first choose $\l \in S^1$ so that the set $\{ \a^l, \; \l \cdot \b^m | 0 \leq l \leq 2, \; 0 \leq m \leq 4 \}$ is symmetric with respect to some axis through the origin, and then choose $\xi$ so that this axis is the imaginary axis.

To finish the proof of Lemma \ref{nonresessential} we show that all the solutions $(k_1, k_2, k_3, k_4)$ of $\sum_{i=1}^4 \sin \frac{k_i \pi}{N} = 0$ obtained in (\ref{sinres1})-(\ref{sinres4}) and the additional ones obtained by replacing $0 < x < \pi$ in $\sin x$ by $\pi - x$ satisfy $\sum_{i=1}^4 k_i \not \equiv 0$ mod $N$ and hence are not in $K_4$.

For the solutions obtained in (\ref{sinres1})-(\ref{sinres4}), $N$ is even. Hence if $N$ is odd, then there is no quadruple $(k_1,k_2,k_3,k_4) \in K_4$ such that (\ref{reseqn}) and (\ref{slsmneq0}) are satisfied. This finishes the proof of Lemma \ref{nonresessential} in this case.

For the rest of the proof, we assume that $N$ is even. If $N=42r$ for some $r \in \N$, (\ref{sinres1}) becomes
\begin{displaymath}
  \sin \frac{7r \pi}{42r} + \sin \frac{9r \pi}{42r} + \sin \frac{(-3r) \pi}{42r} +\sin \frac{(-15r) \pi}{42r} = 0,
\end{displaymath}
and we have $7r+9r-3r-15r = -2r \not \equiv 0$ mod $42r$. Hence the corresponding quadruple $(k_1,k_2,k_3,k_4)$ is not in $K_4$. For the quadruples obtained by replacing $0 < x < \pi$ in $\sin x$ by $\pi - x$ in some of the summands in (\ref{sinres1}), the condition $\sum_{i=1}^4 k_i \not \equiv 0$ mod $42r$ amounts to
\begin{equation} \label{pmeqn42}
  \pm 7 \pm 9 \pm 3 \pm 15 \not \equiv 0 \; \textrm{mod} \; 42
\end{equation}
for any combination of plus and minus signs. The relations (\ref{pmeqn42}) are easily verified. Similarly, one verifies that the quadruples $(k_1,k_2,k_3,k_4)$ satisfying (\ref{sinres2}), (\ref{sinres3}), or (\ref{sinres4}) are not in $K_4$ by showing that
\begin{equation} \label{pmeqn30}
  \pm 5 \pm 13 \pm 7 \pm 9 \not \equiv 0, \quad \pm 5 \pm 1 \pm 11 \pm 3 \not \equiv 0, \quad \pm 15 \pm 5 \pm 3 \pm 9 \not \equiv 0 \; \textrm{mod} \; 30,
\end{equation}
again for any combination of plus and minus signs. Hence we have shown that none of the solutions $(k_1,k_2,k_3,k_4)$ of (\ref{reseqn}) is an element of $K_4$. This completes the proof of Lemma \ref{nonresessential}.
\end{proof}

\begin{proof}[Proof of Lemma \ref{nonreslemmaapp}]
The claimed statement follows from Lemma \ref{nonreslemmaifpart}, \ref{simplecases}, \ref{nonressimple2}, and \ref{nonresessential}.
\end{proof}

\end{document}